\documentclass{llncs}
\usepackage{amssymb}
\usepackage{epsfig}
\usepackage{makeidx}  

\begin{document}

\begin{frontmatter}

\pagestyle{headings}  

\mainmatter

\title{Role of Symmetry and Geometry in a chaotic Pseudo-Random Bit Generator}

\titlerunning{PRBG based on 2D chaotic maps of logistic type}  

\author{Carmen Pellicer-Lostao \and Ricardo L\'{o}pez-Ruiz}

\authorrunning{Pellicer-Lostao and L\'{o}pez-Ruiz  }   

\institute{Department of Computer Science and BIFI, \\
Universidad de Zaragoza, 50009 - Zaragoza, Spain,\\
\email{carmen.pellicer@unizar.es,}
\email{rilopez@unizar.es} }

\maketitle              

\begin{abstract}
In this work, Pseudo-Random Bit Generation (PRBG) based on 2D chaotic mappings of logistic type is considered.
The sequences generated with two Pseudorandom Bit Generators (PRBGs) of this type are statistically
tested and the computational effectiveness of the generators is estimated. 
The role played by the symmetry and the geometrical properties of the underlying chaotic attractors 
is also explored. Considering these PRBGs valid for
cryptography, the size of the available key spaces are calculated. Additionally, a novel mechanism called
\emph{symmetry-swap} is introduced in order to enhance the PRBG algorithm. It is shown that it can increase the
degrees of freedom of the key space, while maintaining the speed and performance in the PRBG.

\end{abstract}

\end{frontmatter}

\section{Introduction}

Pseudo-Random Bit (or Number) Generation is a subject of high interest and broad application in many scientific
and engineering areas ~\cite{nieder},~\cite{anderson},~\cite{menezes}. Pseudo-Random Bit Generators (PRBGs) are
implemented by deterministic numeric algorithms and they should pass several statistical tests
~\cite{nist},~\cite{marsaglia},~\cite{knuth}, to prove themselves to be useful. The requirements of randomness
in these generators vary according to their application, realizing in cryptography their most stringent demands
~\cite{nist}.

Over the last two decades, several works have implemented PRBGs for cryptography based on chaotic systems (an
extensive survey can be found in~\cite{Li_thesis}). Chaotic systems have the property of being deterministic in
the microscopic space and behave randomly, when observed in a coarse-grained state-space. Their sensitivity of
chaotic maps to initial conditions make them optimum candidates to relate minimal critical information about the
input in the output sequence ~\cite{protopopescu}. Their iterative nature makes them fast computable and able to
produce binary sequences with extremely long cycle lengths ~\cite{kocarev2}.

In 2006, Madhekar Suneel proposes in ~\cite{suneel} a method for pseudo-random binary sequence generation based
on the two-dimensional H\'{e}non map. The  pseudorandom sequences generated with this algorithm show good random
properties when subjected to different statistical tests suites. The author also indicates that the choice of
the H\'{e}non map is rather arbitrary and that similar results should also be attainable with other 2D maps.

The present explores precisely this possibility, and presents a finite automata scheme as the key to achieve
that. This comprehensive scheme is then applied to two particular chaotic maps presented in ~\cite{ric}. These
2D dynamical systems are formed by two symmetrically coupled logistic maps. The refined knowledge of the chaotic
systems under study (i.e., its geometry) makes possible to obtain the finite automata and to extent the method
in ~\cite{suneel} to this type of chaotic mappings. The pseudo-random properties of the generators obtained that
way are investigated. The evaluation of the potential range of input parameters and the computational cost of
the algorithm makes them worth to be considered for cryptographic applications.

The chaotic PRBG algorithm here described can be used in different ways. Focusing on cryptography, one of its
applications and maybe the most immediate, could be the construction of practical stream ciphers. In this way,
the chaotic PRBG can expand a short key into a long keystream, which directly exclusive-or'ed with the clear
text or message, gives the ciphertext.

The interest of PRBG based on these mappings of logistic type arises from the fact that they present interesting
geometrical symmetries. These could offer additional advantages to the randomization algorithm. In fact, a novel
mechanism to enhance PRBG is proposed in this paper. This mechanism could be applied to chaotic PRBGs based on
mappings with the same symmetry characteristics.

The paper is structured as follows: Section 2 introduces some basic concepts about chaotic PRBG. Section 3
describes statistical testing to asses PRBG randomness. Section 4 explains the PRBG algorithm applied to the
H\`{e}non map and infers the finite automata that describes its dynamics. In Section 5 the finite automata scheme is
used to obtain PRBG based on a two-dimensional symmetrical chaotic map of logistic type. Several sequences are
obtained and their randomness is tested. The computational cost and key space for cryptographic applications are
effectively evaluated. Section 6 presents an enhancement of the PRBG algorithm, based on the symmetry properties
of these chaotic maps. Section 7 exposes the final conclusions.

\section{Chaotic Random Bit Generation}

The inherent properties of chaos, such as ergodicity and sensitivity to initial conditions and control
parameters, connect it directly with cryptography characteristics of confusion and diffusion
~\cite{alvarez},~\cite{Li_degrad}.

Additionally chaotic dynamical systems have the advantage of providing simple computable deterministic
pseudo-randomness. As a consequence of these observations, several works were presented since 1990s implementing
PRBGs based on different chaotic systems ~\cite{protopopescu}, ~\cite{kocarev2}, ~\cite{suneel}, ~\cite{Lee},
~\cite{Li_CCS}. In some way, it could be said that chaos has brought into being a novel branch of PRBGs in
cryptography, called chaotic PRBGs.

An N-dimensional deterministic discrete-time dynamical system is an iterative map $f:\Re^N\rightarrow\Re^N$   of
the form:

\begin{equation}
X_{k+1} = f (X_k)
\end{equation}

where $k = 0,1\ldots n$. is the discrete time and $X_0,X_1\ldots X_n$, are the states of the system at different
instants of time.

In these systems, the evolution is perfectly determined by the mapping $f:\Re^N\rightarrow\Re^N$ and the initial
condition $X_{0}$. Starting from $X_{0}$, the \textit{initial state}, the repeated iteration of (1) gives rise
to a fully deterministic series of states known as an \textit{orbit}. Different models of $N-$dimensional
discrete-time mappings have been studied, and under certain circumstances complex behaviour in time evolution
has been shown. The one-dimensional cases have been deeper analyzed ~\cite{collet}, the cases with N=2 have also
several well explored examples ~\cite{mira}, but as N increases the complexity grows and less literature is
found with a well documented analysis of the chaotic properties of the mapping ~\cite{ric2}.

To build a chaotic PRBG is necessary to construct a numerical algorithm that transforms the states of the system
in chaotic regime into binary numbers. The existing designs of chaotic PRBGs use different techniques to pass
from the continuum to the binary world ~\cite{Li_thesis}. The most important are:

\begin{enumerate}
    \item Extracting one or more bits from each state along chaotic orbits ~\cite{protopopescu},~\cite{pascal}.
    \item Dividing the phase space into m sub-spaces, and output a binary number $i={0,1,…,m-1}$ if
     the chaotic orbit visits the $i_{th}$ subspace
     ~\cite{kocarev2}, ~\cite{suneel}.
    \item Combining the outputs of two or more chaotic systems to generate the pseudo-random numbers
     ~\cite{Lee},~\cite{Li_CCS}.
\end{enumerate}

At that point an important divergence appears. We have to remark that chaos implemented on computers with finite
precision is normally called``pseudo chaos". In pseudo chaos dynamical degradation of the chaotic properties of
the system may appear, for throughout iterations pseudo orbits may depart from the real ones in many different
and uncontrolled manners ~\cite{Li_thesis},~\cite{shadow}.

Even so, the above exposed techniques are capable of generating sequences of bits, which appear random-like from
many aspects. One must only consider their implementation in a sensitive way to minimize dynamical degradation.
Therefore a detailed study is normally required on the dynamics of the chaotic system. This will guarantee that
the PRBG passes the required statistical tests and can be easily implemented with simple and fast software
routines.

As a final hint to help in this process, one may consider as an advantage the idea of using high dimensional
chaotic systems. While less known, these systems whirl many variables at any calculation. Therefore the periodic
patterns produced by the finite precision of the computer are more difficult to appear than in the low
dimensional case ~\cite{vulpiani}.

In this paper the technique of dividing the phase space is followed and applied on two symmetrical
two-dimensional (2D) chaotic maps of logistic type.

\section{Statistical Tests Suites}

In general, randomness cannot be mathematically proved. Alternatively, different statistical batteries of tests
are used. Each of these tests evaluates a relevant random property expected in a true random generator. These
properties may correspond to specific physical systems or to given statistical characteristics. To test a
certain randomness property, several output sequences of the generator under test are taken. As one knows a
priori the statistical distribution of possible values that true random sequences would be likely to exhibit for
that property, a conclusion can be obtained upon the probability of the tested sequences to be random.

Mathematically this is done as follows ~\cite{nist}. For each test, a \emph{statistic variable} $X$ is specified
along with its correspondent \emph{theoretical random distribution function} $f(x)$. For non-random sequences,
the statistic can be expected to take on larger values, typically far-out in the tails of $f(x)$. A
\emph{critical value} $x_{\alpha}$ is defined for the theoretical distribution so that $P(X>x_{\alpha})=
\alpha$, that is called the \emph{significance level} of the test. In the same way, theoretically other
distribution functions and a $\beta$ value could be defined to assess non-random properties. But in practice, it
is impossible to calculate all distributions that describe non-randomness, for there are an infinite number of
ways that a data stream can be non-random.

When a test is applied, the test statistic value $X_s$ is computed on the sequence being tested. This test
statistic value $X_s$ is compared to the critical value $x_{\alpha}$. If the test statistic value exceeds the
critical value, the hypothesis for randomness is rejected. The rejection is done with a $(100*\alpha)\%$
probability of having FALSE POSITIVE error. This is called a \emph{TYPE I error}, where the sequence was random
and is rejected. Otherwise is not rejected (i.e., the hypothesis is accepted) with a probability of
$(100*\beta)\%$ of error. This is called \emph{TYPE II error} or FALSE NEGATIVE, the sequence was non-random and
is accepted. As a consequence, passing the test merely provides a probabilistic evidence that the generator
produces sequences which have certain characteristics of randomness.

For a given application, the value of $\alpha$ must be selected appropriately. This is because if $\alpha$ is
too high, TYPE I errors may frequently occur (respectively, if $\alpha$ is too low the same will happen for TYPE
II errors). For cryptographic applications typical values of $\alpha$ are selected in the interval
$\alpha\epsilon[0.001, 0.01]$, which is also referred as a confidence level for the test in the interval
$[99.9\%, 99\%]$. Unlike $\alpha, \beta$ is not fixed, for it depends on the non-randomness defects of the
generator. Nevertheless $\alpha,\beta$ and the size ($n$) of the tested sequence  are related. Then for a given
statistic, a critical value and a minimum $n$ should be selected to minimize the probability of a TYPE II error
$(\beta)$.

There exist different well-known sources of test suites available, such as those described by Knuth
~\cite{knuth}, the Marsaglia´s Diehard test suite ~\cite{marsaglia} or those of the National Institute of
Standards and Technology (NIST) ~\cite{nist}. But there are many more, perhaps not so nicely packaged as in the
works mentioned above, but still useful (~\cite{rutti}, ~\cite{mascagni}, etc.) . In these collections of tests,
each test tries a different random property and gives a way of interpreting its results.

In the present work, Marsaglia's Diehard test suite (in ~\cite{marsaglia}) and NIST Statistical Test Suite (in
~\cite{nist}) were selected, for they are very accessible and widely used. Table 1 lists the tests comprised in
these suites.

\begin{table} [h]
\label{table1}
\begin{center}
\begin{tabular}{|c|c|c|}
   \hline
  Number & Diehard test suite & NIST test suite \\
  \hline
  1 & Birthday spacings & Frequency (monobit) \\
  2 & Overlapping 5-permutation  & Frequency test within a block \\
  3 & Binary rank test &  Cumulative sums\\
  4 & Bitstream & Runs\\
  5 & OPSO & Longest run of ones in a block\\
  6 & OQSO & Binary matrix rank\\
  7 & DNS & Discrete fourier transform\\
  8 & Count-the-1's test & Non-overlapping template matching\\
  9 & A parking lot & Overlapping template matching\\
  10 & Minimum distance & Maurer's universal statistical\\
  11 & 3D-spheres & Approximate entropy\\
  12 & Squeeze &  Random excursions\\
  13 & Overlapping sums & Random excursions variant\\
  14 & Runs & Serial\\
  15 & Craps & Linear complexity\\
  \hline
\end{tabular}
\end{center}
\caption{List of tests comprised in the Diehard and NIST test suites.}
\end{table}

In each test, the statistic value $X_s$ is obtained and used to calculate a p-value that summaries the strength
of evidence against the randomness of the tested sequence. In Marsaglia's Diehard test suite, p-values should
lie within the interval $[0,1)$ to accept the PRBG. In NIST Statistical Test Suite, p-values should be greater
than $\alpha$ for acceptance.

\section{Pseudo-Random Bit Generation based on the H\`{e}non Map}

In ~\cite{suneel}, an algorithm is presented to obtain a chaotic PRBG using the H\'{e}non map. The H\'{e}non map
~\cite{henon} is a 2D discrete-time non linear dynamical system represented by the state equations:

\begin{equation}
\begin{array}{l}
x_{k+1} = ax_k^2 + y_k + 1,\\
y_{k+1} = bx_k.
\end{array}
\end{equation}

This system depends on two parameters, a and b. Depending on the values of these parameters the system may be
chaotic, intermittent, or converge to a periodic orbit. The map has a so called canonical form for the parameter
values $a=1.4$ and $b=0.3$ which is depicted in Fig. \ref{fig2}. For the canonical values the H\'{e}non map presents
a chaotic attractor. This means that an initial point of the plane will either approach a set of points known as
the H\'{e}non strange attractor, or diverge to infinity.

\begin{figure}[]
\centerline{\includegraphics[width=12cm]{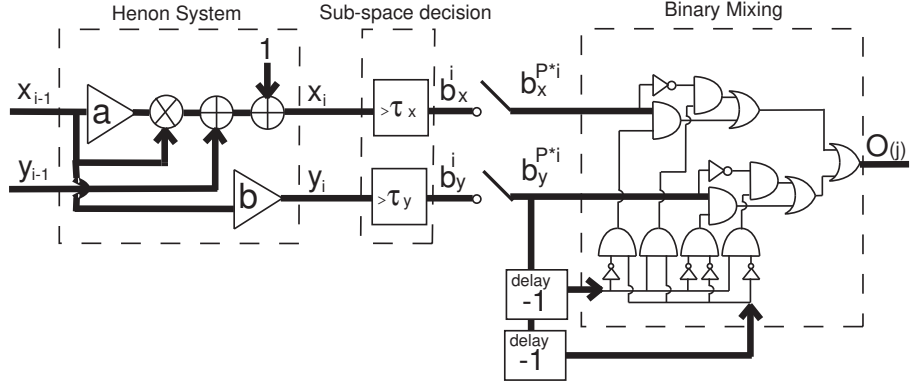}} 
\caption{Functional structure of M. Suneel's numerical algorithm.} 
\label{fig1}
\end{figure}

\begin{figure}[h]
\centerline{\includegraphics[width=6.35cm]{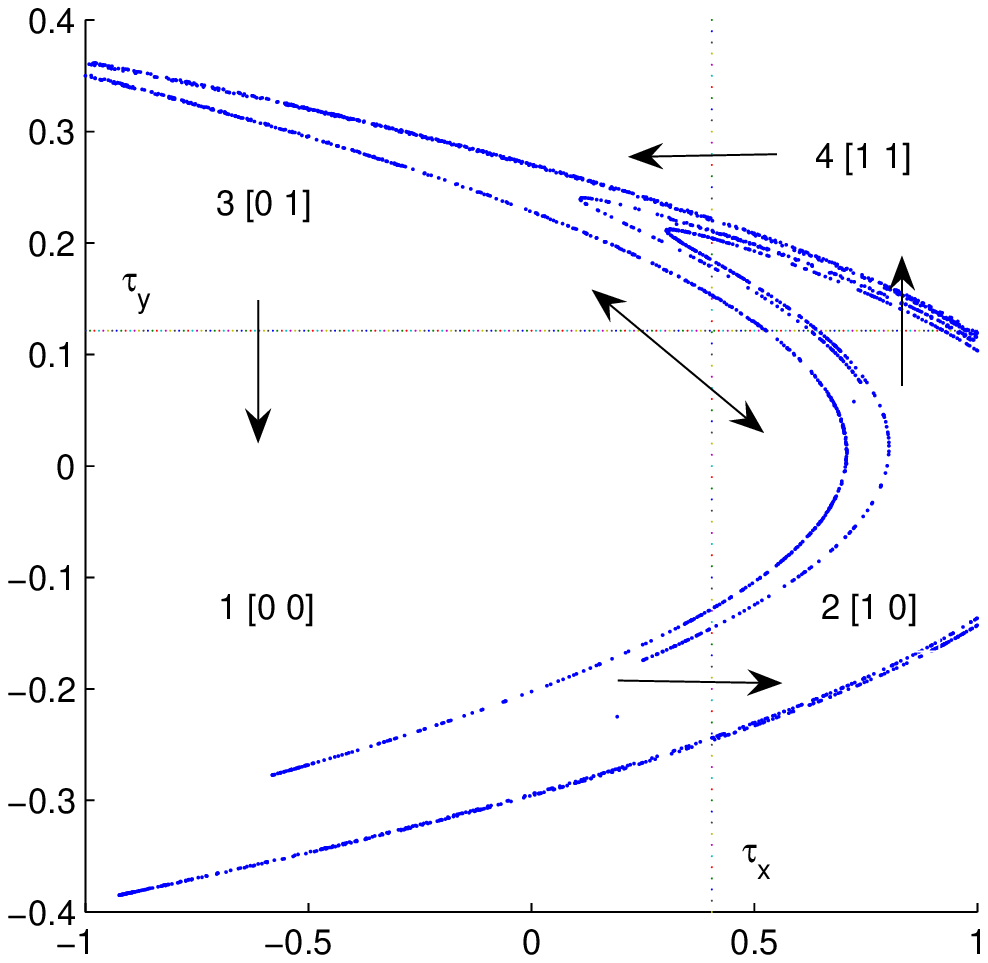}\hskip 6mm\includegraphics[width=4.3cm]{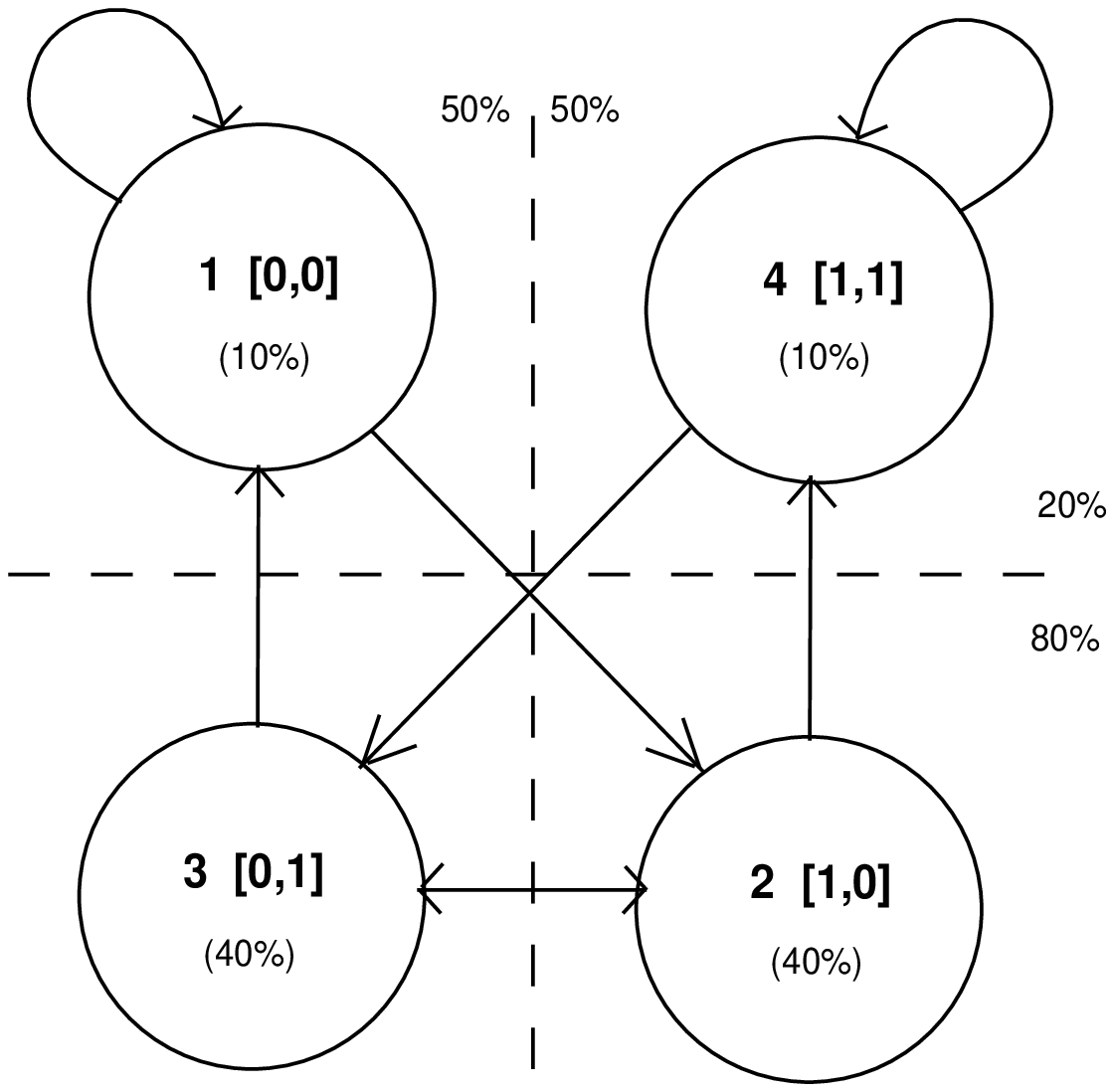}}
\centerline{(a)\hskip 5.5cm (b)} \caption{(a) Representation of the canonical H\`{e}non map with parameters values
$a=1.4$ and $b=0.3$. The picture shows 3000 iterations from the initial state $X_0=[-0.75,-0.02]$. The $\tau$
values calculated after the first $T=1000$ iterations are $\tau_x=0.404659$ and $\tau_y=0.121397$.(b) Finite
state automaton summarizing the distribution of visit of each sub-space.} 
\label{fig2}
\end{figure}

In Fig. \ref{fig1} the functional block structure of algorithm ~\cite{suneel} is represented and it is explained
in the following paragraphs.

In this case, the technique of dividing the phase space in four sub-spaces is used. This is done in the block
named as \textit{Sub-space decision} in which the threshold values, $\tau_x$ and $\tau_y$, are employed to
convert the points into a binary sequence, by means of the following equations:

\begin{equation}
b_x = \left\lbrace
\begin{array}{l}
0 \hskip 2mm {\rm if} \hskip 2mm x  \hskip 2mm \leq \hskip 2mm  \tau_x \\ 1  \hskip 2mm {\rm if} \hskip 2mm x
\hskip 2mm
> \hskip 2mm \tau_x
\end{array};
\right.
 \hskip 2cm
b_y =  \left\lbrace
\begin{array}{l}
0 \hskip 2mm {\rm if} \hskip 2mm y  \hskip 2mm \leq  \hskip 2mm \tau_y \\ 1  \hskip 2mm {\rm if} \hskip 2mm y
\hskip 2mm > \hskip 2mm \tau_y
\end{array}.
\right.
\end{equation}

A purely statistical procedure is proposed in ~\cite{suneel} to obtain $\tau_x$ and $\tau_y$. They are
calculated as the medians of a large $T$ set of $x$ values (for $\tau_x$) and $y$ values (for $\tau_y$). More
precisely, the value of $\tau_x$ and $\tau_y$ are the medians of the first $T=1000$ iterations of the system.
Fig. \ref{fig2}(a) shows, as an example, one orbit of the H\'{e}non map with the $\tau$ values and subspaces
considered for that case.

After obtaining $S_x=\{b_x^i\}_{i=1}^{\infty}$ and $S_y=\{b_y^i\}_{i=1}^{\infty}$, they are sampled with a
frequency of $1/P$ (each $P$ iterations) and $B_x=\{b_x^{P*i} \}_{i=1}^{\infty}$ and
$B_y=\{b_y^{P*i}\}_{i=1}^{\infty}$ are obtained. The effect of skipping P consecutive values of the orbit is
necessary to get a random macroscopic behaviour. With this operation, the correlation existing between
consecutive values generated by the chaotic system is eliminated, in a way such that over a $P_{min}$, sequences
generated with $P>P_{min}$ will appear macroscopically random. Although $P$ is normally introduced as an
additional key parameter in pseudo-random sequences generation ~\cite{kocarev3}, it strongly determines the
speed of the generation algorithm. Consequently it is recommended to be kept as small as possible.

The output binary pseudorandom sequence $O(j)$ is obtained in the block named \textit{Binary mixing} in
Fig. \ref{fig1}. Here a mixing operation is performed with the current and previous values of the sequence
$B(j)=[B_x(j),B_y(j)]$. The operation is given by the truth table sketched in Table 2.

\begin{table}
\label{table2}
\begin{center}
\begin{tabular}{|c|c|c|}
  \cline{2-3}
  \multicolumn{1}{c|}{} & \multicolumn{2}{|c|}{$B_y(j-1)$}\\
   \hline
  $B_y(j-2)$ & 0 & 1\\
  \hline
   0 & $B_x(j)$ & ${\rm Not}(B_x(j))$\\
   \hline
   1 & $B_y(j)$ & ${\rm Not}(B_y(j))$\\
  \hline
\end{tabular}
\end{center}
\caption{Truth table generating the binary sequence.}
\end{table}



In the exposed algorithm the selection of the $\tau$ values is the determinant factor for a uniform distribution
of each of the coordinates of the phase states within different sub-spaces. According to ~\cite{suneel}, these
values should be chosen in a way that approximately half of the $x$ (or $y$ values) obtained over the iterations
of the system lay at each side of the threshold.

This fact leads us to consider the interest of analyzing in detail the operations performed in the \textit{H\`{e}non
system} and \textit{Sub-space decision} blocks of Fig. \ref{fig1}. The objective is to trace the visits of the
orbit states into each sub-space, consequently gaining knowledge of how to obtain the binary sequence $[b_x^i,
b_y^i]$.

Fig. \ref{fig2}(a) presents the evolution of the \textit{H\`{e}non system} in the phase space for a given set
parameters and initial conditions. The phase space is divided in 4 sub-spaces, which are named as 1,2,3 and 4
according with the different outputs $[b_x^i, b_y^i]$ of the \textit{Sub-space decision} block. The output
values $[0,0]$, $[1,0]$, $[0,1]$ and $[1,1]$ correspond to sub-spaces names 1,2,3 and 4 respectively.

According to Fig. \ref{fig2}(a), at one instant of time $i$, one sub-space is visited and this give the
corresponding values $[b_x^i,b_y^i]$. As the system evolves, the movement between the different sub-spaces can
be resumed in the finite state automaton represented of Fig. \ref{fig2}(b). This automata summarizes the complete
behaviour of the blocks under study and can be described as follows:

\begin{description}
    \item[(a)] The $80$\% of the time, there is a bi-directional oscillation between sub-spaces 2 and 3 (equally balanced with a $50$\%).
Apparently there are no consecutive visits of the same sub-space. This leads to an oscillation of $[1,0]$ and
$[0,1]$ between the binary states of the sequence $[b_x^i,b_y^i]$.
    \item[(b)] The rest $20$\% of the visits are equally distributed into sub-spaces 1 and 4 (with a $50$\% each). It may make small runs in 1 or
4, but normally the system mainly spins in counter clock direction around the center of the subspaces division.
It circles around 3-1-2, or 2-4-3 to fall in the 3-2 oscillation, or it makes a complete round along 1-2-4-3.
This leads to an oscillation between the binary states of the sequence $b_y^i$ (entry value in binary mixing
equation of Table 2, as $B_y(j-1)$ and $B_y(j-2)$).
\end{description}

Although the four sub-spaces are not visited equally, there exists a symmetry of movements between sub-spaces
1-3 and 2-4, which has a characteristic mixing of 50\% and 50\%, as long as a predominant (80\%) and constant
transition between 3 and 2. This leads to a highly variation of binary values in sequences $S_x$, $S_y$. In the
end, these conditions give the final result of an output sequence $O(j)$ with a proper balance of zeros and
ones, or put it in another way, with pseudo-random properties.

\section{Pseudo-Random Bit Generation based on two-dimensional chaotic maps of logistic type}

As it has been exposed the choice of the H\'{e}non in Fig. \ref{fig1} was rather arbitrary and similar results
should also be attainable with other 2D maps ~\cite{suneel}. This means that a substitution of the \textit{H\`{e}non
system} block by any other 2D chaotic system would potentially produce $O(j)$ sequences with pseudo-random
properties.

To prove the generality of the algorithm, the present paper explores the application of Fig. \ref{fig1} to a
specific family of 2D chaotic maps, quite different in nature and geometrical properties to the H\`{e}non map.


\subsection{Pseudo-Random Bit Generator}

In ~\cite{ric}, L\'{o}pez-Ruiz and P\'{e}rez-Garc\'{\i}a analyze a family of three chaotic systems obtained by coupling two
logistic maps. The focus here will be made on models (a) and (b), which will be called Logistic Bimap system A
and B:

\begin{equation}
\label{systems}
\begin{array}{l}
SYSTEM\;\; A:\\ T_A:[0,1]\times[0,1]\longrightarrow[0,1]\times[0,1]\\ \; \\
x_{n+1} =\lambda(3y_n+1)x_n(1-x_n)
\\ y_{n+1}=\lambda(3x_n+1)y_n(1-y_n)
\end{array}
\hskip 0.5 cm
\begin{array}{l}
 SYSTEM\;\; B:\\ T_B:[0,1]\times[0,1]\longrightarrow[0,1]\times[0,1]\\ \; \\
 x_{n+1} =\lambda(3x_n+1)y_n(1-y_n)
 \\ y_{n+1} =\lambda(3y_n+1)x_n(1-x_n)
\end{array}
\end{equation}

Amazingly, these systems show the following symmetry $T_A(x,y)=T_B(y,x)$, which implies that
$T_A^2(x,y)=T_B^2(x,y)$. From a geometrical point of view, both present the same chaotic attractor in the
interval $\lambda\in[1.032, 1.0843 ]$. The dynamics in this regime is particularly interlaced around the saddle
point $P4$, that plays an important role for our proposes:

\begin{equation}
\label{p4} P4=[P4_x, P4_y],\hskip 8mm  where \hskip 4mmP4_x=P4_y=\frac{1}{3}
\left(1+\sqrt{4-\frac{3}{\lambda}}\right).
\end{equation}

\begin{figure}[h]
\centerline{\includegraphics[width=5.2cm]{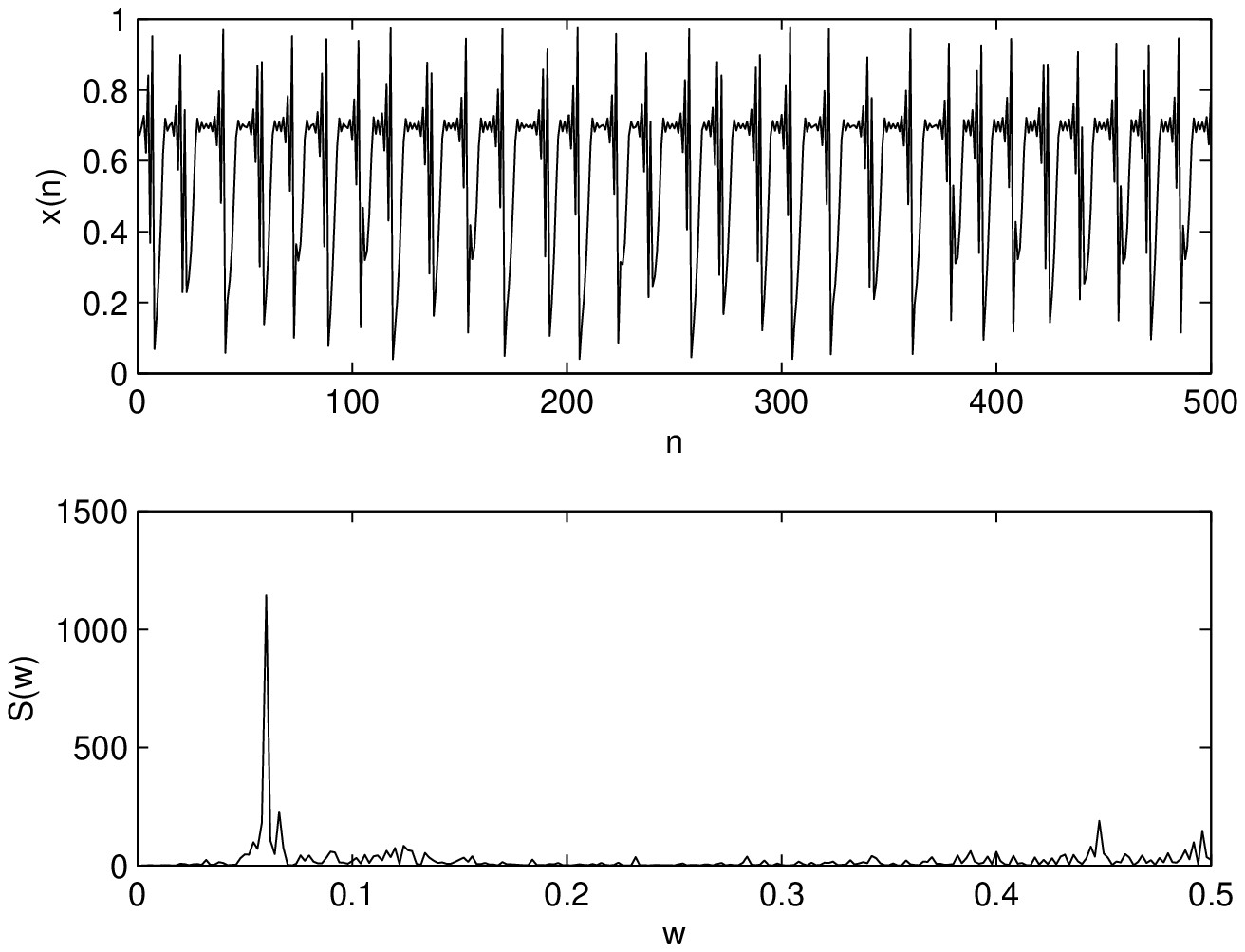}\hskip 5mm\includegraphics[width=5.2cm]{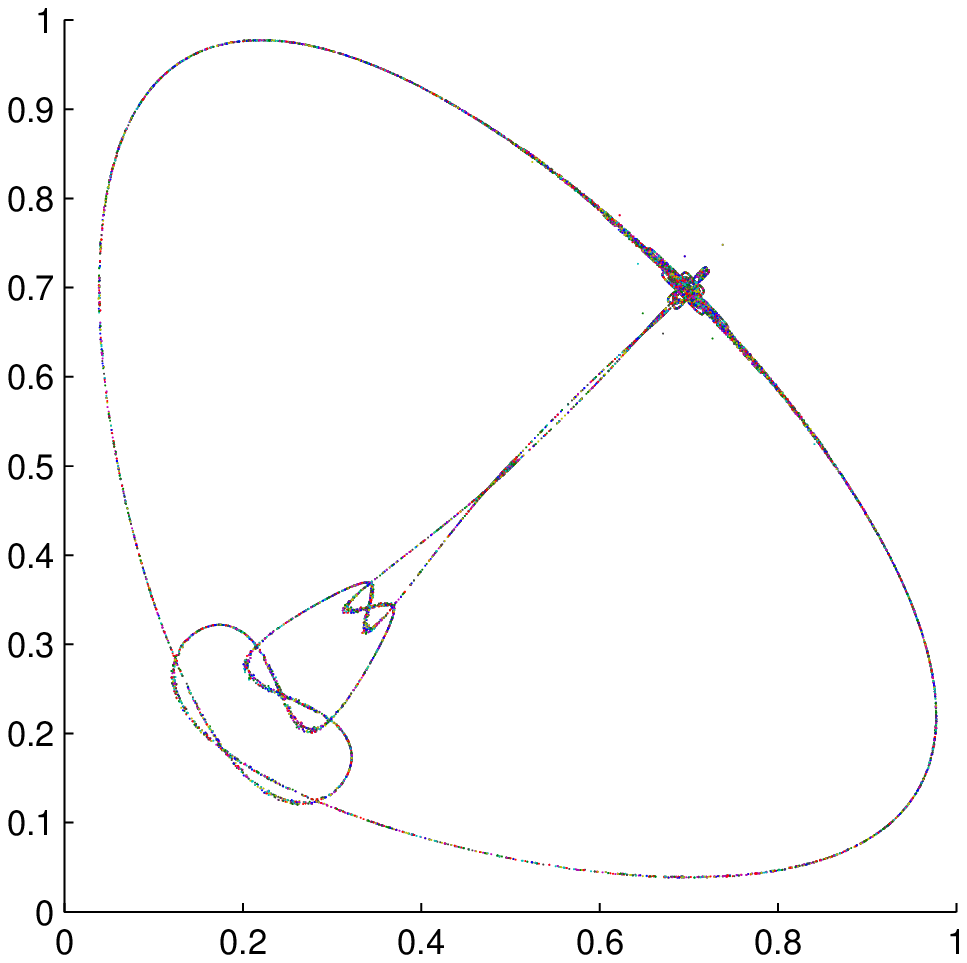}}
\centerline{(a)\hskip 5cm (b)} \centerline{\includegraphics[width=5.2cm]{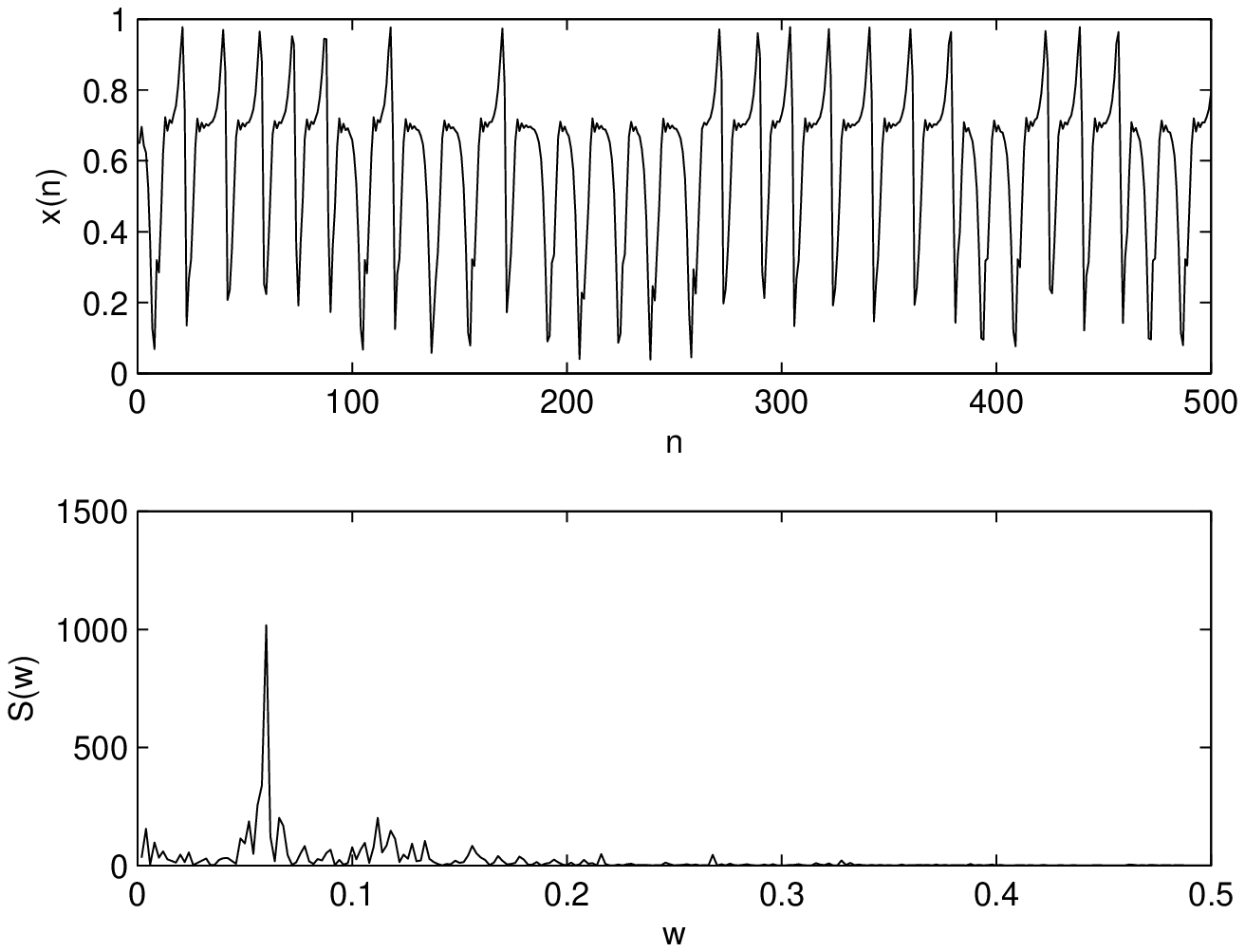}\hskip
5mm\includegraphics[width=5.2cm]{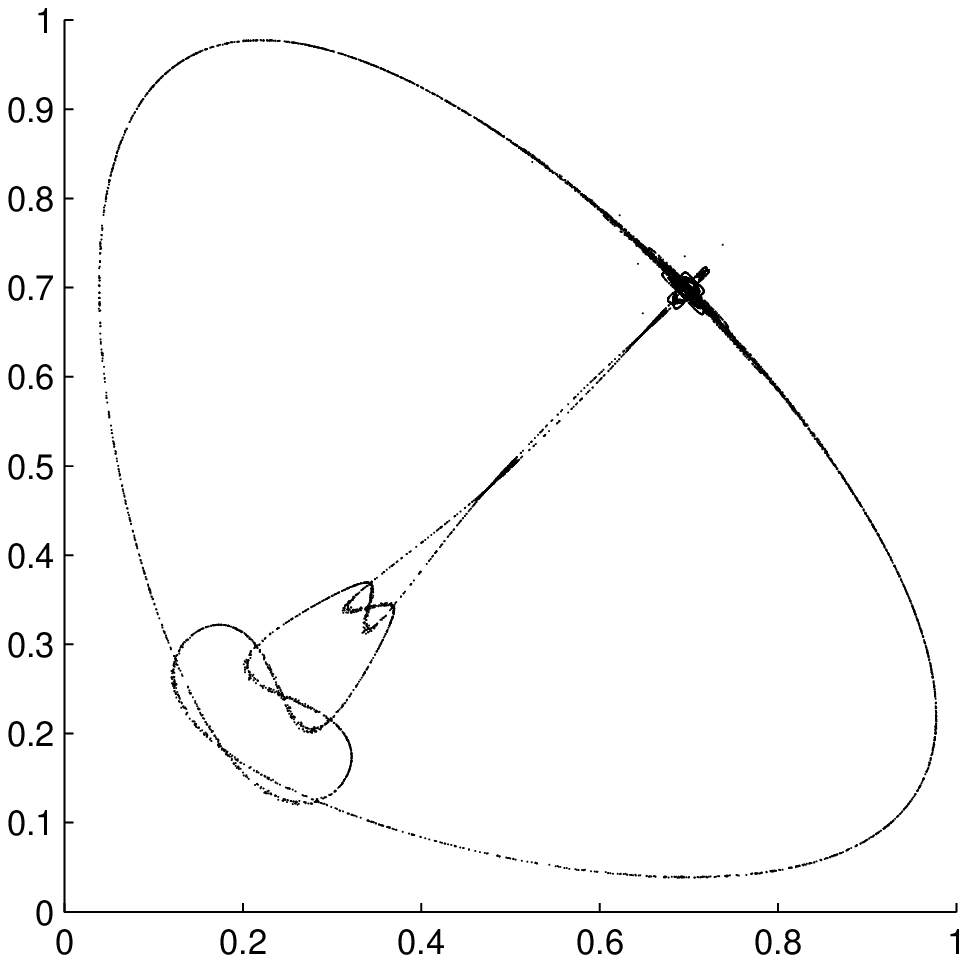}} \centerline{(c)\hskip 5cm (d)} \caption{Chaotic attractor, temporal
iterations and spectrum with $\lambda= 1.07$ and $X_0=[0.737, 0.747]$ for System A ((a),(b) respectively) and
System B ((c),(d) respectively).} 
\label{fig3}
\end{figure}

On the other hand, their dynamics have some differences. In Fig. \ref{fig3} one can see one orbit and the
spectrum for both systems with a given set of equal initial conditions.

It can be seen in Fig. \ref{fig3} (a) and (c), that the movements in the orbits have two lobules at each side of
the diagonal axis, folding in $P4$, where the dynamics of the systems turns out to become erratic. The spectrum
of this movement is also shown in Fig. \ref{fig3} (b) and (d), where the differences can be appreciated. System
A produces an oscillation of period two, that makes it jump over the diagonal axis alternatively between points
consecutive in time.

\begin{figure}[]
\centerline{\includegraphics[width=12cm]{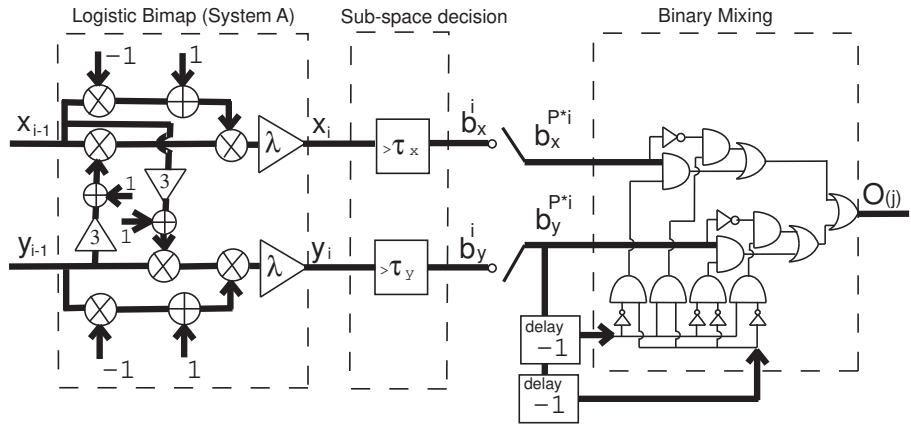}} 
\caption{Functional block structure of the proposed algorithm
in \cite{suneel} for System A.} 
\label{fig1b}
\end{figure}

To obtain the Symmetric Coupled Logistic Map PRBG, the algorithm presented in ~\cite{suneel} is applied on
System A. Its functional block structure is represented in Fig. \ref{fig1b}.  The threshold values  $\tau_x$ and
$\tau_y$ are calculated as the medians of $x_i$ and $y_i$ values calculated for a large set of $T=1000$ initial
iterations.

Different $O(j)$ binary sequences are created with Fig. \ref{fig1b} and submitted to statistical testing as
described in section 3. Unfortunately the sequences so formed do not pass the minimum requirements of randomness
assessed by Diehard Test Suite. The results are found to get worse for larger shift values $P$ or longer
sequences. Similar results were obtained for System B. Therefore algorithm in Fig. \ref{fig1} it is no directly
applicable to other 2D chaotic maps. Something else must be taken into consideration in this approach.

At this point, it was found that the merely equal-statistical division of $x_i$ and $y_i$ components by median
threshold values for a number of initial iterations does not work. Moreover it was found that, one must select
the division lines between sub-spaces so that the new 2D chaotic system follows a finite state automata similar
to the one depicted in Fig. \ref{fig2}(b). Consequently, the geometrical characteristics of the system must be
taken into account.

That means that the substitution of the \textit{H\`{e}non system} block is not enough to extend the algorithm to
other 2D chaotic systems. One needs also replace the \textit{sub-space decision} block. Therefore a refined
knowledge of the geometrical properties of the chaotic system is a priory required to build the PRBG. This makes
the extension of the algorithm possible in fact, but not so straight forward. The knowledge of the necessary
finite automata can help to make the process systematic.

Let us apply the finite automata scheme to the symmetric coupled logistic maps Systems A and B. To get this
automata one should chose the diagonal axis, as the first division line. This is because, this axis divides
phase space in two parts each of which is equally visited (50\%). And additional statistical calculus is
required to divide these two sub-spaces, in another two with a visiting rate of 40\%-10\% each one.

When this is done, one can observe that this is got by merely selecting P4 and the line perpendicular to the
diagonal axis in P4 as the other division line. This gives the final sub-space division independence of the
initial point or iterations and a different geometrical division from the initial cartesian proposal. As a last
step,  the sub-spaces are finally labeled (1,2,3 or 4) according with their position in the finite automata to
match \ref{fig2}(b).

\begin{figure}[h]
\centerline{\includegraphics[width=4cm]{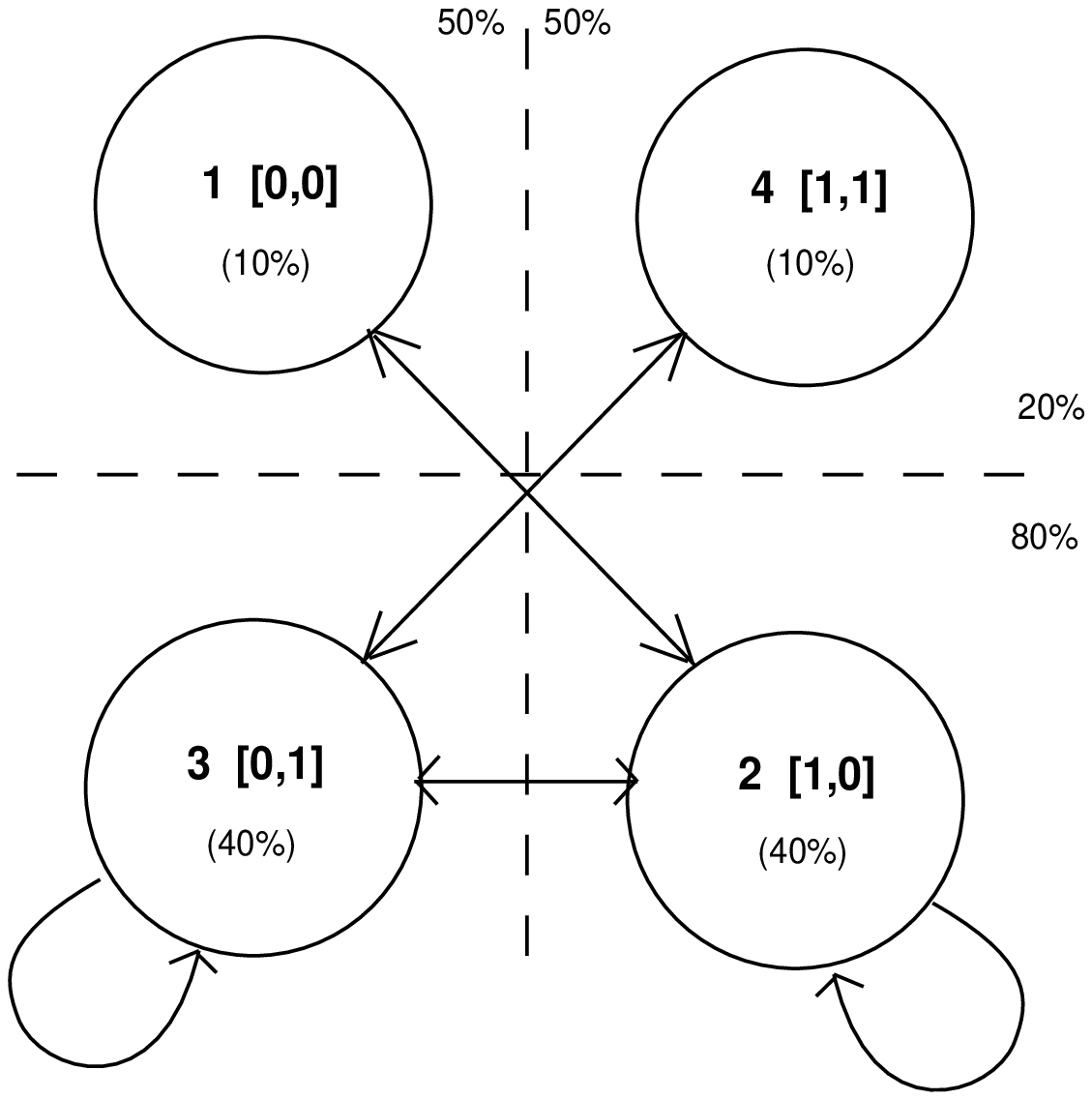}\hskip 5mm\includegraphics[width=5.8cm]{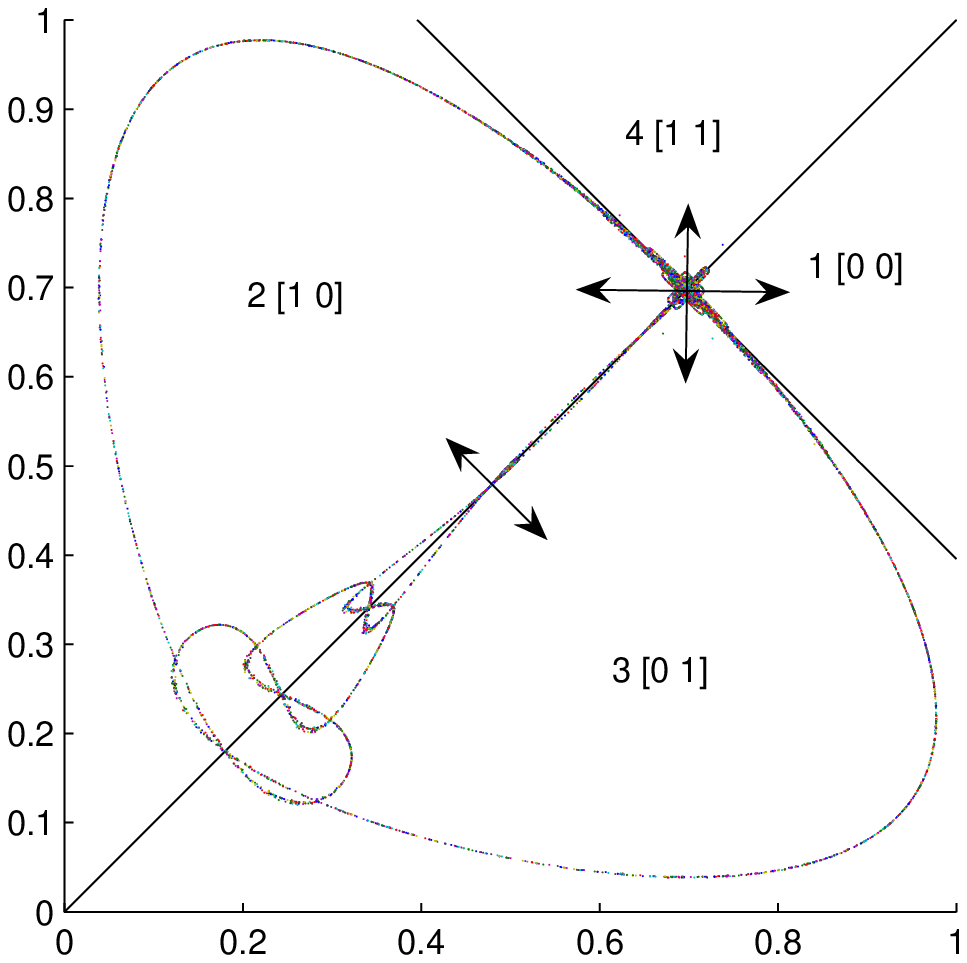}}
\centerline{(a)\hskip 5.3cm (b)} \centerline{\includegraphics[width=4cm]{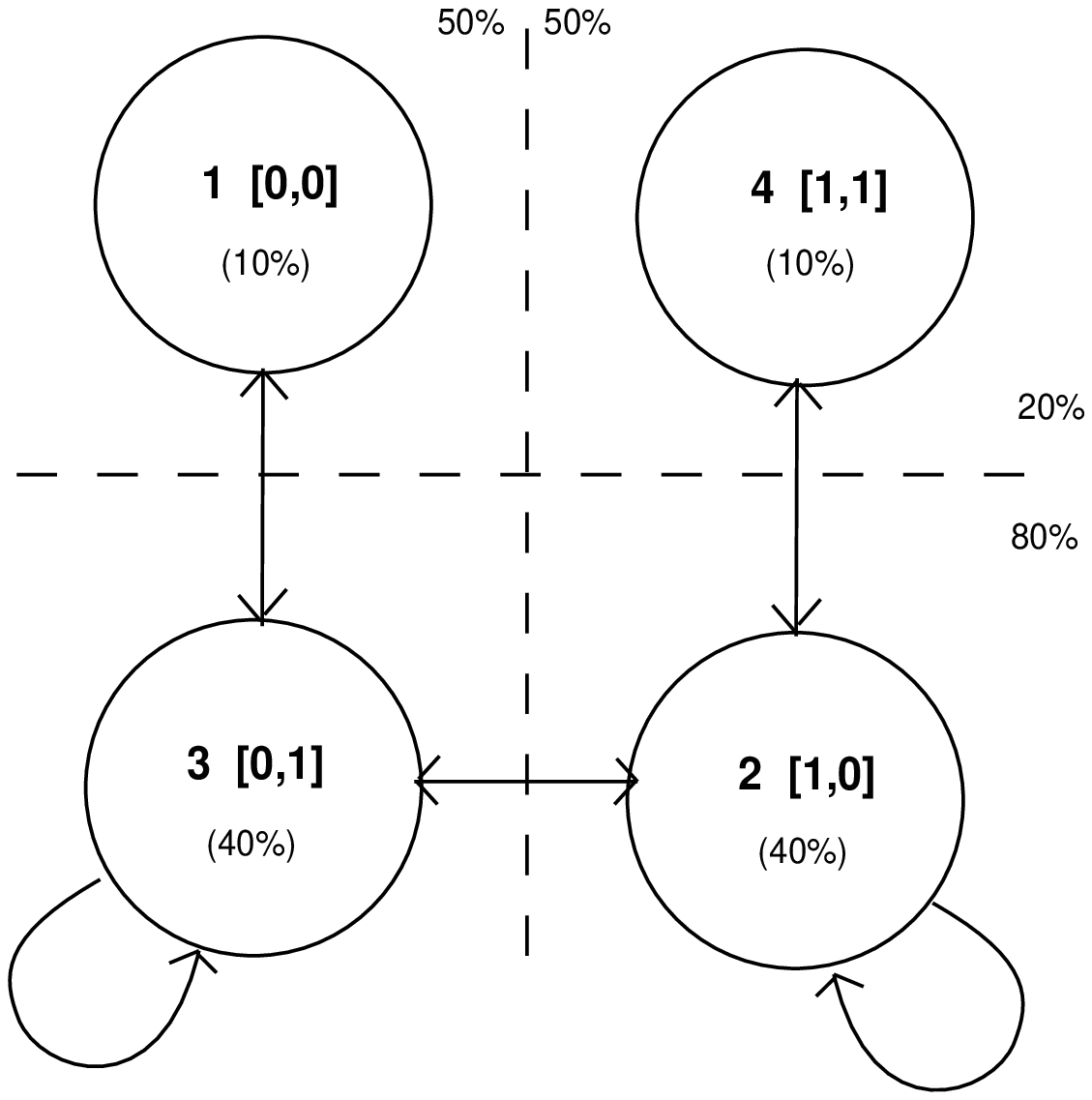}\hskip
5mm\includegraphics[width=5.8cm]{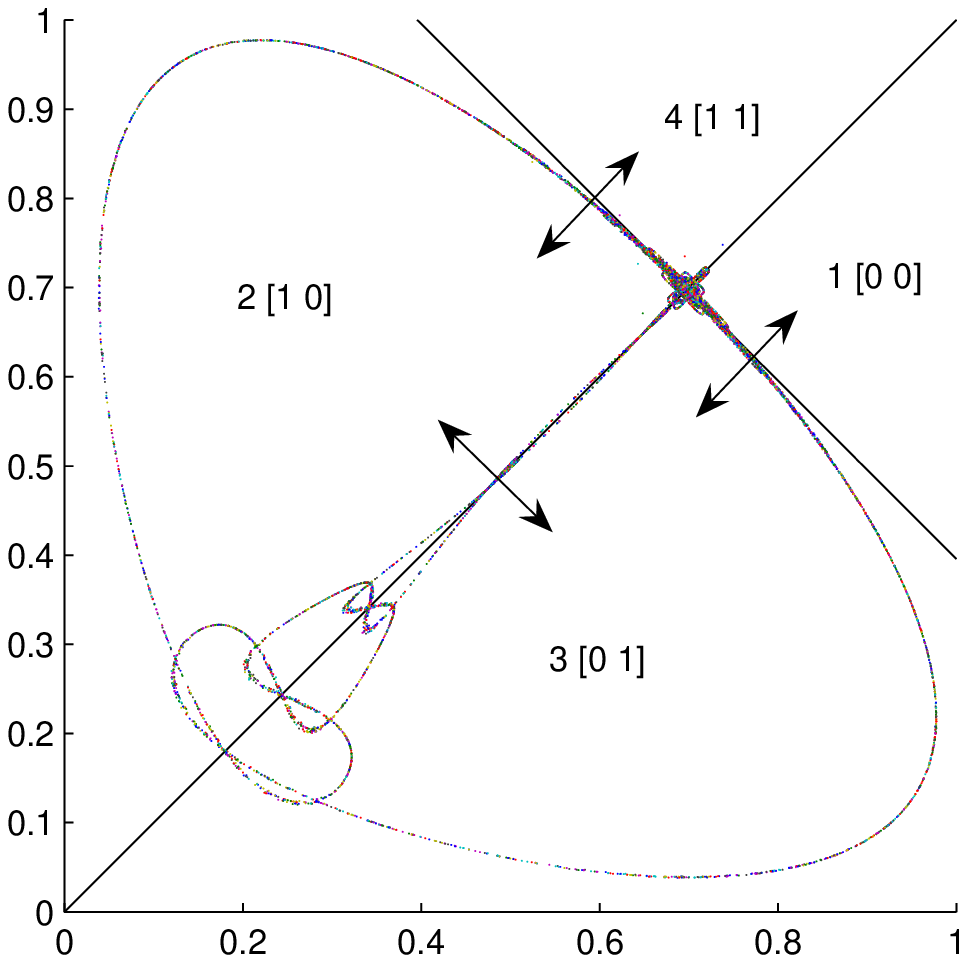}} \centerline{(c)\hskip 5.3cm (d)} 
\caption{(a) Finite automata and (b) final sub-space division for System A. 
(c) Finite automata and (d) final sub-space division for System B. (In
both cases, $\lambda= 1.07$).} 
\label{fig4}
\end{figure}

The final sub-space division for each system is presented in Fig. \ref{fig4}(b) and \ref{fig4}(d), along with
the indications of the evolution of the visits to each sub-space.

Both systems posses similar statistical properties with different movement across the diagonal axis. The
automata represented in Fig. \ref{fig4} (a) and (c), and that of Fig. \ref{fig2}(b) are similar in many aspects.
The only difference between them is the pace of consecutive visits take place, but the mixing proportions of
50\%-50\% and 80\%-20\% are maintained. From this finite automata is possible to build the required
\textit{Sub-space decision} block.

\begin{figure}[h]
\centerline{\includegraphics[width=12cm]{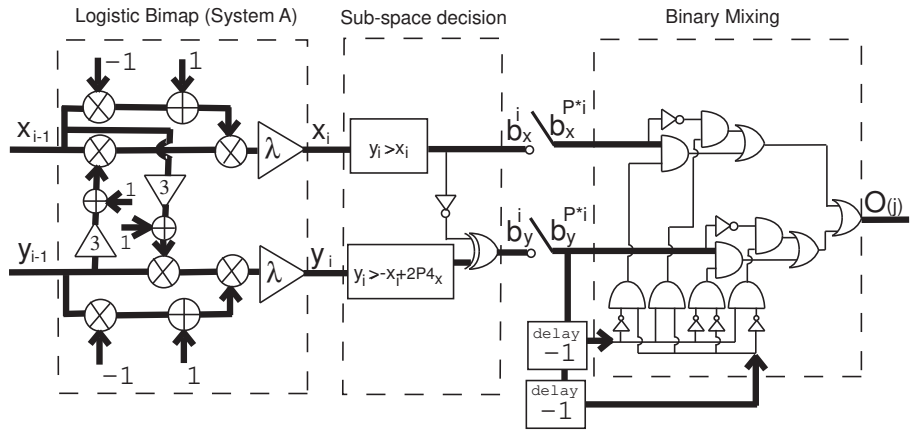}} 
\caption{Functional block structure of the PRBG applied to the
symmetric coupled logistic map PRBG with System A.} 
\label{fig6}
\end{figure}

Finally the initial algorithm in Fig. \ref{fig1} applied to System A, is modified with the appropriate
\textit{Sub-space decision} block. The final PRBG functional scheme is represented in Fig. \ref{fig6}.

Different sequences are obtained with the system of Fig. \ref{fig6} in next sub-section. Their randomness is
assessed and it demonstrates them statistically valid for cryptographic applications. This may indicate that the
automata scheme described here represents a sufficient condition to obtain pseudo-randomness.

Consequently, it may represent a systematic scheme to extend the algorithm in ~\cite{suneel} to get PRBG on
other chaotic maps. The cost of this algorithm and its hypothetical achievable key-space for cryptographic
applications are also estimated in subsection 5.3.

\subsection{Pseudo-Random Sequences and Statistical testing}

To assess the randomness of the PRBG obtained in the previous section with systems A and B, several sequences
are obtained and submitted to the Diehard ~\cite{marsaglia} and NIST ~\cite{nist} test suites described in
section 3. The significance level of the tests was set to a value appropiate for cryptographic applications
($\alpha=0.01$).

Similar results were found for both systems and for simplicity, only those obtained with system A will be
presented here after. Ten sequences were generated with six different sets of initial conditions. Their
characteristics are described in Table 3.

\begin{table} [h]
\label{table3}
\begin{center}
\begin{tabular}{|c|c|c|c|c|c|c|}
  \hline
  Sequence & S1 & S2 & S3 & S4 & S5 & S6\\
  \hline
  $x_0$ & $0.989125$ & $0.491335$ & $0.672757$ & $0.726874$ & $0.39565$ & $0.999851$\\
  $y_0$ & $0.689125$ & $0.691335$ & $0.497757$ & $0.901874$ & $0.49565$ & $0.649851$\\
  $\lambda$ & $1.04869$ & $1.05392$ & $1.06961$ & $1.08007$ & $1.06438$ & $1.07489$\\
  $P_{Dmin}$& $55$ & $45$ & $35$ & $47$ & $n.a.$ & $n.a.$ \\
  $P_{Nmin}$& $83$ & $105$ & $83$ & $83$ & $100$ & $85$ \\
  \hline
\end{tabular}
\end{center}
\caption{Parameters $P_{Dmin}$ and $P_{Nmin}$ for different sequences $S_i$, $i=1,..,6$, with different initial
conditions $(x_0,y_0)$ and map parameter $\lambda$.}
\end{table}

Six of them (S1,S2,S3,S4,S5 and S6) were tested with Nist tests suite with 200 Mill. of bits and four of them
(S1,S2,S3 and S4) were tested with Diehard tests suite with 80 Mill. of bits. Here, the parameters $P_{Dmin}$
and $P_{Nmin}$ are the minimum sampling rate or shift factor, $P_{min}$, over which, all sequences generated
with the same initial conditions and $P>P_{min}$ pass Diehard or Nist tests suites, respectively. It is observed
here, that the Nist tests suite requires a higher value of $P_{min}$ and that S5 and S6 were not tested with
Diehard battery of tests.

In the Diehard tests suite, each of the tests returns one or several p-values which should be uniform in the
interval [0,1) when the input sequence contains truly independent random bits. The software available in
~\cite{marsaglia} provides a total of 218 p-values for 15 tests, and the uniformity requirement can be assessed
graphically, when plotting them in the interval [0,1).

For example Fig. \ref{fig7b} shows the p-values obtained for three sequences (a),(b) and (c) with the same
initial conditions S1, and different sampling factor P.

\begin{figure}[h]
\centerline{\includegraphics[width=6.3cm]{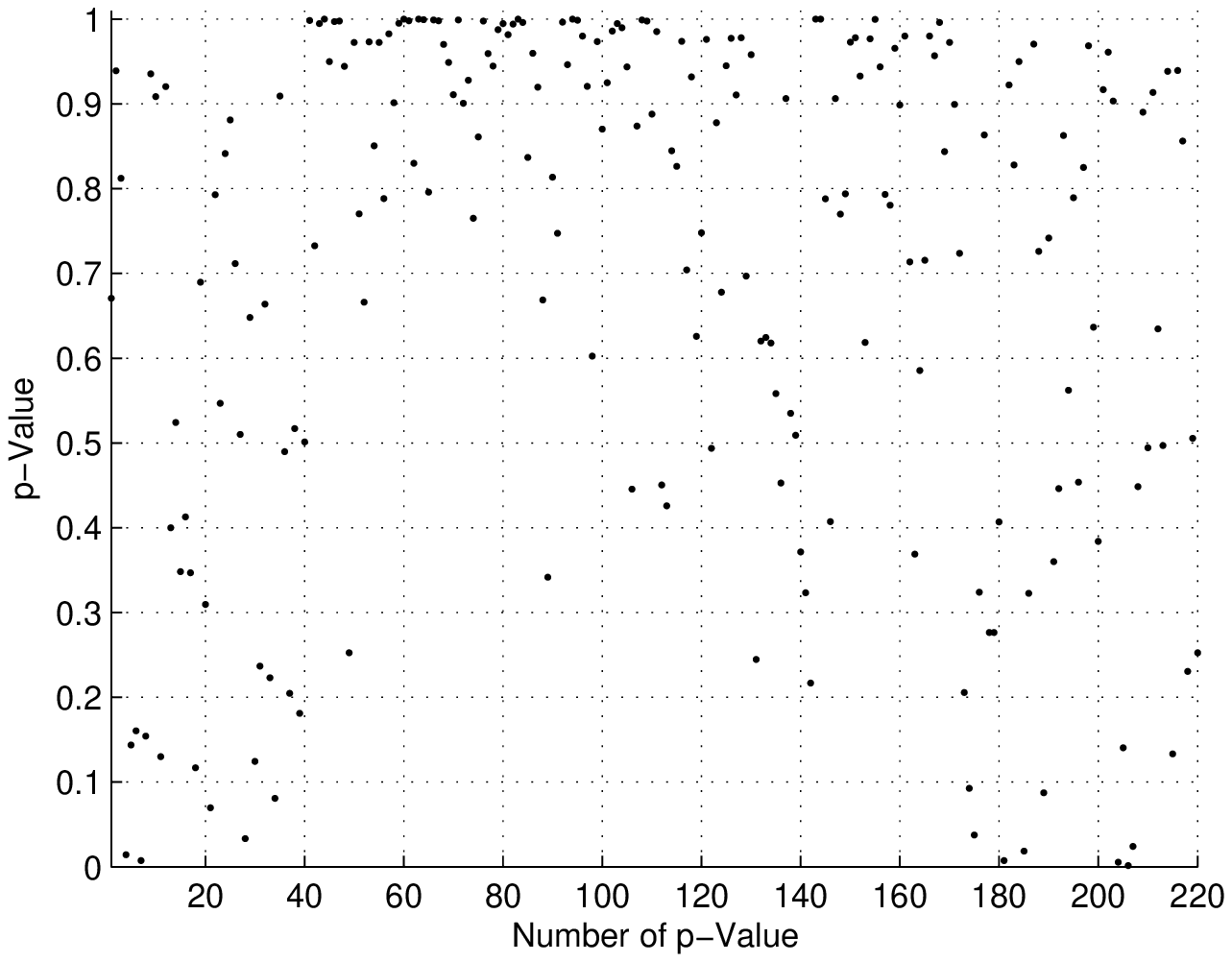}\hskip 5mm\includegraphics[width=6.3cm]{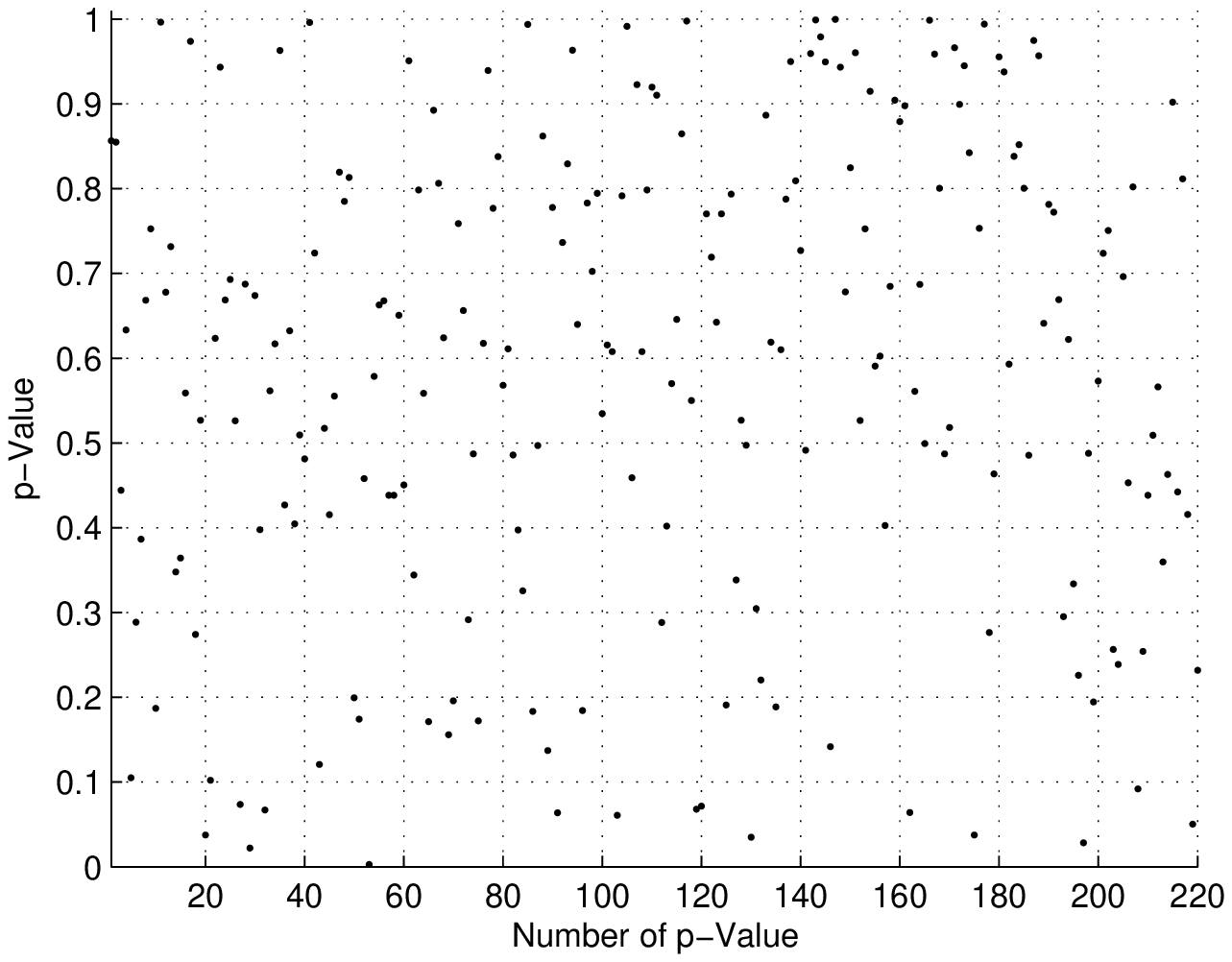}}
\centerline{(a)\hskip 6cm (b)}
 \label{fig7a}
\end{figure}

\begin{figure}[h]
\centerline {\includegraphics[width=6.3cm]{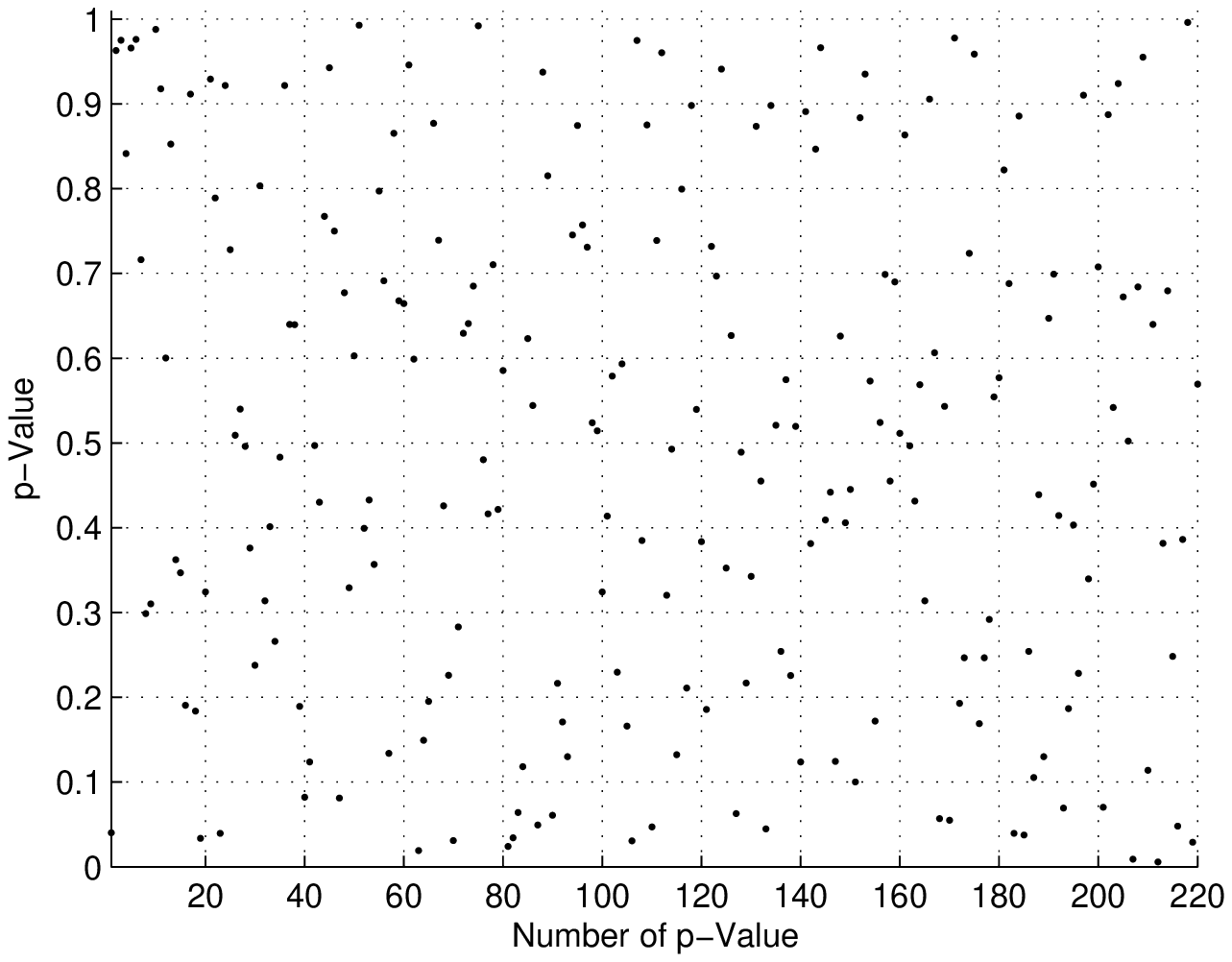}\hskip 5mm\includegraphics[width=6.3cm]{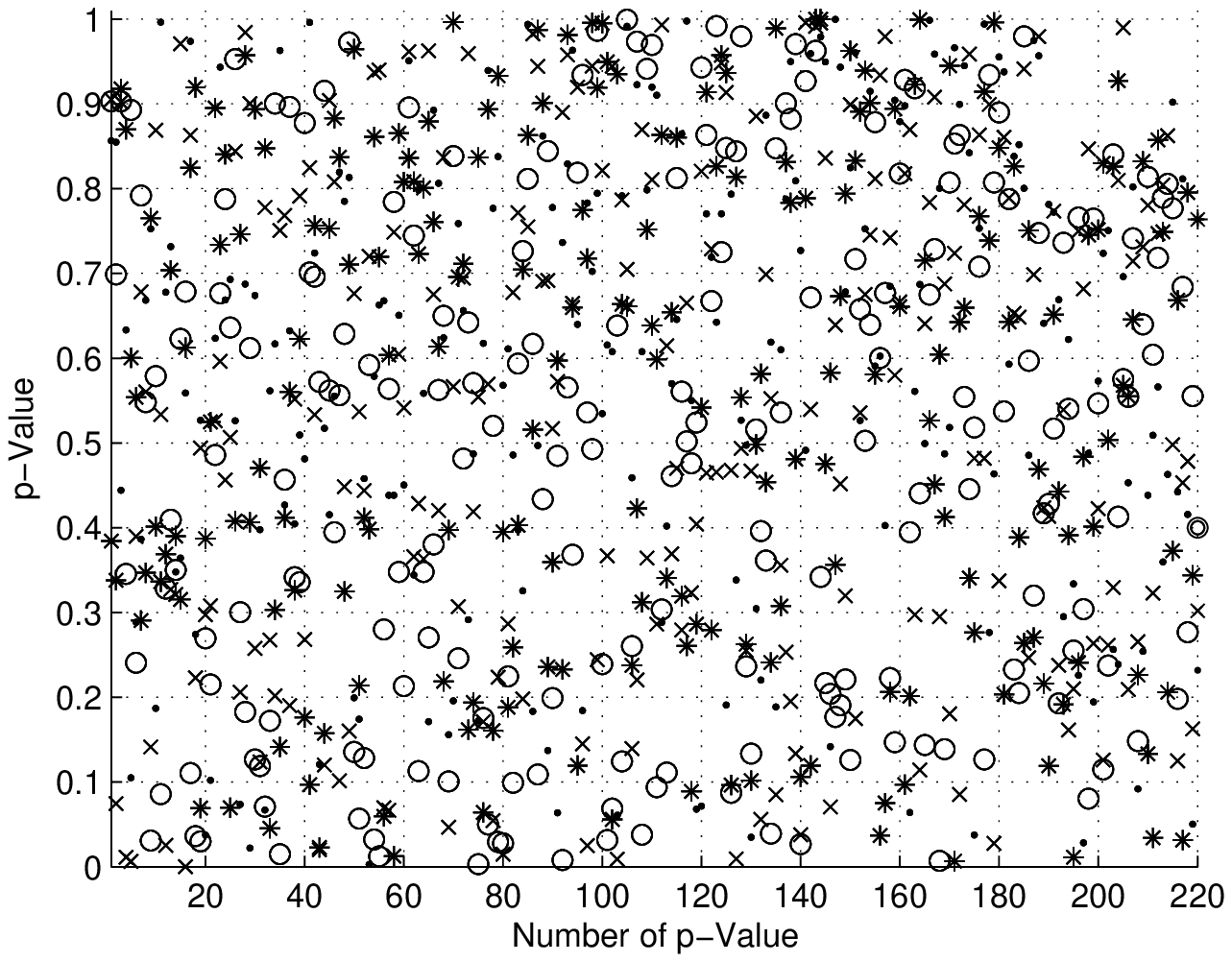}}
\centerline{(c)\hskip 6cm (d)}
\caption{Diehard test suite p-values obtained with all tests for initial
conditions S1 with (a) $P=20$ , (b) $P=P_{Dmin}=55$ and (c) $P=110$ . 
In (d), p-values obtained for initial
conditions S1,S2,S3 and S4 with $P=P_{Dmin}$ of Table 3.} 
\label{fig7b}
\end{figure}

The first one, Fig. \ref{fig7b}(a), demonstrates graphically the failure of the tests, for there is a
non-uniform clustering of p-values around one. Fig. \ref{fig7b}(b) shows the uniformity obtained with
$P_{Dmin}=55$ over the interval [0,1). A better uniformity can be appreciated when $P>P_{Dmin}$ in Fig.
\ref{fig7b}(c).

 Sequences S1 to S4 where proved to pass the Diehard battery of tests with significance level
$\alpha=0.01$. Fig.\ref{fig7b}(d) presents a graphical representation of the p-values obtained for each sequence
with sampling factor $P=P_{Dmin}$ of Table 3. It can be observed that some p-values are occasionally near 0 or
1. Although it can not be well appreciated in the figure, it has to be said that those never really reach these
values.

In the Nist tests suite ~\cite{nist}, one or more p-values are also returned for each sequence under test. These
values should be greater than the significance level $\alpha$, which was selected to $\alpha=0.01$ as in the
Diehard case. These tests also require a sufficiently high length of sequences and to prove randomness in one
test, two conditions should be verified. First, a minimum percentage of sequences should pass the test and
second, the p-values of all sequences should also be uniformly distributed in the interval $(0,1)$.

For this case, each of the six sequences with initial conditions S1 to S6 are arranged in 200 sub-sequences of
1Mill. bits each and submitted to the Nist battery of tests. Sequences $S$ proved to pass all tests over a
minimum value $P_{Nmin}$, shown in Table 3.

\begin{figure}[h]
\centerline{\includegraphics[width=6cm]{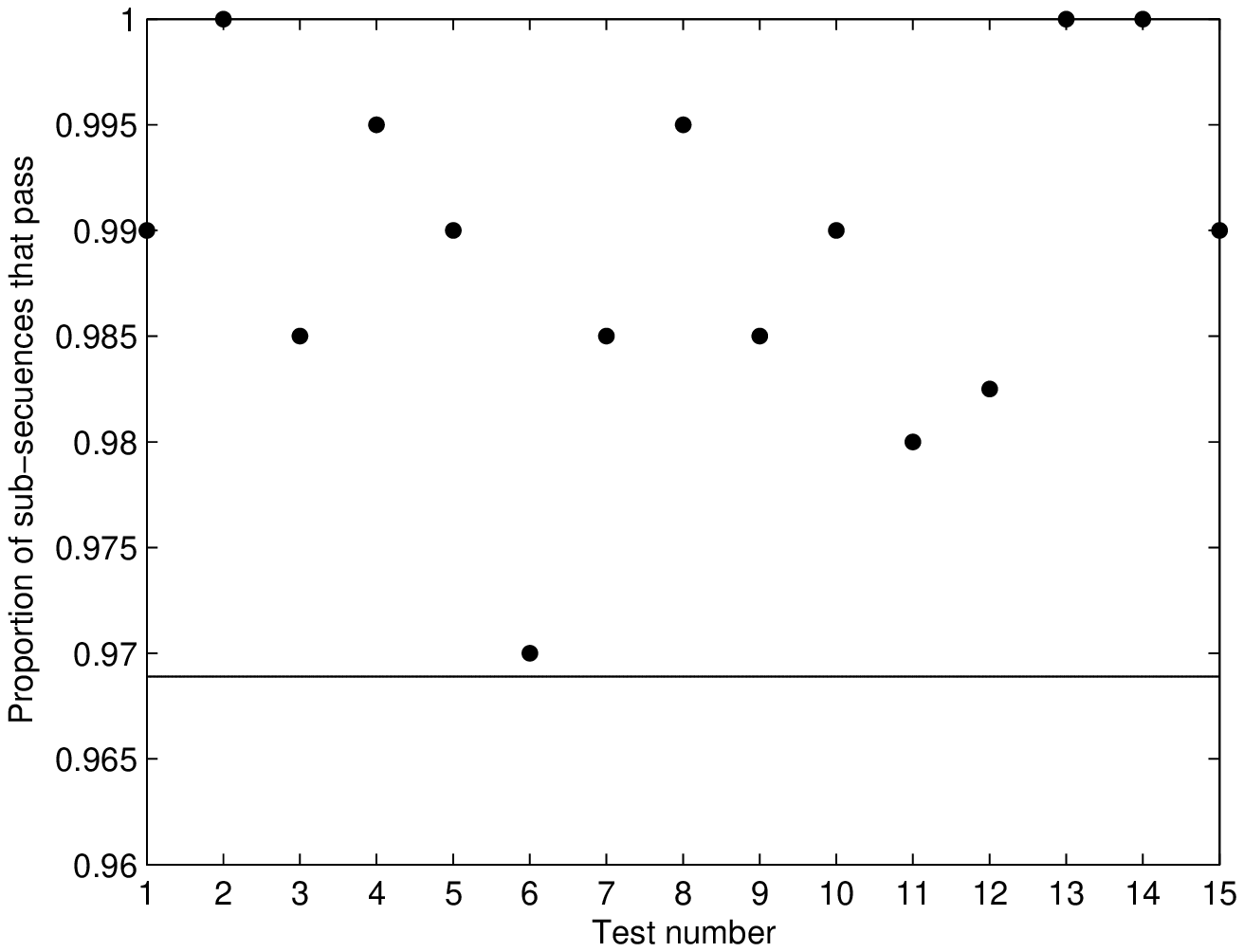}\hskip 0mm\includegraphics[width=6.5cm]{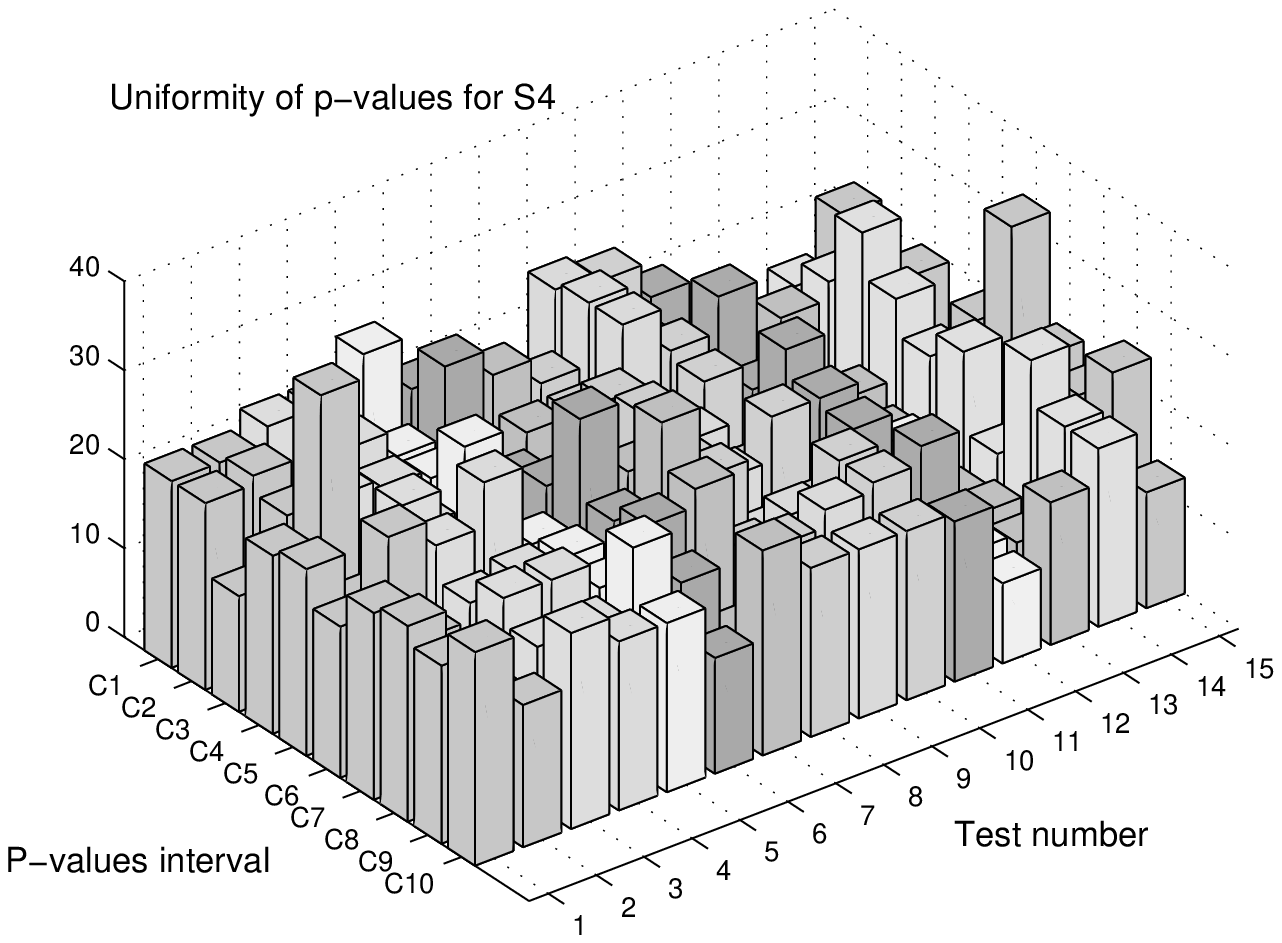}}
\centerline{(a)\hskip 5.5cm (b)}\caption{In (a), the proportion of sub-sequences of S1 that passes each test is
displayed. In (b) The distribution of p-values of S4 is examined for each test to ensure uniformity. The
interval between 0 and 1 is divided in ten sub-intervals $(C1,C2,...,C10)$, and the p-values that lie within
each subinterval are counted and plotted.} 
\label{fig8}
\end{figure}

In Fig.\ref{fig8}(a) and \ref{fig8}(b), the results obtained for $S1$ and $S4$ respectively are graphically
presented, as an example of what was obtained for each $S$. The tests in the suite are numbered according to
Table 1. Fig. \ref{fig8}(a) represents the percentage of the 200 sub-sequences of S1, that pass each of the 15
tests of the suite. These percentages are over the minimum pass rate required of $96.8893\%$ for a sample size =
200 binary sub-sequences. Fig. \ref{fig8}(b) describes the uniformity of the distribution of p-values obtained
for the 15 tests of the suite. Here, uniformity is assessed. The interval (0,1) is divided in ten subintervals
$(C1,C2,...,C10)$ and the number of p-values that lay in each sub-interval, among a total of 200, are counted
and proved to be uniform.

\subsection{Key space size and computational cost}

To establish the complexity, and consequently the speed of the PRBG described in Fig. \ref{fig6}, the principle
of invariance is observed. This says that the efficiency of one algorithm in different execution environments
differs only in a multiplicative constant, when the values of the parameters of cost are sufficiently high.

In this sense, the asymptotic behaviour of the computational cost of the PRBG is governed by the calculus
performed in the chaotic block. This block performs P iterations to obtain an output bit, $O(j)$.

The capital theta notation ($\Theta$) can be used to describe an asymptotic tight bound for the magnitude of
cost of the PRBG. And consequently, the 2D symmetric coupled logistic maps have a computational cost or
complexity of order $\Theta(P*n)$.

Let us determine the operative range of initial conditions and parameters values that can be applied to the PRBG
in Fig. \ref{fig6}. This range, when the PRBG is used in cryptography applications is known as the key-space.
Then, this range or the key space is determined by the interval of the parameter $\lambda$ and the initial
conditions that keep the dynamical system in the chaotic regime. These are $\lambda\in[1.032, 1.0843 ]$ ,
$x_0\in(0, 1)$ and $y_0\in(0, 1)$. The sampling parameter can also be considered as another parameter of the key
space. One must observe that $P$ should be kept in a suitable range, so that the PRBG is fast enough for its
desired application.

These intervals can be denoted with brackets and calculated as $[\lambda]=0.0523$, $[x_0]=1$, $[y_0]=1$ and
$[P]=8890$, when taking $[P]\in[110, 9000]$ as the range of the sampling factor.

Let us consider $\epsilon_{32}\thickapprox1.1921\times10^{-7}$ as the smallest available precision for
fixed-point representation with 32 bits and its correspondent magnitude
$\epsilon_{64}\thickapprox2.2204\times10^{-16}$ for floating-point numbers with 64 bits. These quantities give
us the maximum number of possible values of every parameter in any of the two representations. This is easily
computed dividing the intervals by $\epsilon$, as $K_{\lambda}=[\lambda]/\epsilon$, $K_{x_0}=[x_0]/\epsilon$,
$K_{y_0}=[y_0]/\epsilon$ and $K_{P}=[P]/\epsilon$.

The total size of representable parameter values is given by $K$, calculated as $K= K_{\lambda}\times
K_{x_0}\times K_{y_0}\times K_{P}$. $K$ is the size of the available key-space and its logarithm in base 2 gives
us the available length of binary keys or entries to produce pseudo-random sequences in the generator.

The values obtained for each number precision, are $K_{32}= 2.32\times10^{30}$ with a key length of 100 bits for
single precision and $K_{64}= 1.91\times10^{65}$, with a key length of 216 bits for double precision. These
results are encouraging for recommending the use of the PRBG in Fig. \ref{fig6} for cryptographic applications,
where a length of keys greater than 100 is considered strong enough against brute force attacks,
~\cite{alvarez}.

Nevertheless, it has to be said in the sake of accuracy that the calculus of the key space is a coarse
estimation and that a deeper study is required for an exact evaluation ~\cite{alvarez}. One must keep in mind
that chaotic systems are highly sensitive to the parameter values as well as to the initial conditions and a
slightly change in its values can produce very different evolutions, even taking the system from a quasi-random
behaviour to a periodic orbit. This can be easily understood if one thinks of the chaotic attractor as an
infinite conglomerate of orbits which are periodic and unstable. This means that the system jumps from one to
another without stabilizing in any of them.

This is the origin of its instability and of its apparent macroscopic random behaviour. With a minimum change in
the parameters or initial conditions an immense quantity of bifurcations are taking place. This means that many
periodic orbits are being created and others are disappearing. So it is possible that apparently valid
contiguous values in the key-space lead to periodic and random behaviour respectively in each case. These
phenomena is even more exaggerated when computational precision is taken into account. The continuum chaotic
trajectories are truncated and periodicity is prone to appear with more intensity. Another possibility is that
the dynamics can diverge towards infinity. In the systems presented here an initial calculus of 100 iterations
is enough to ensure the boundless or goodness of the initial conditions.

\section{A new PRBG based on chaotic variables swapping}

The PRBG obtained in Fig.\ref{fig6} on the Symmetric Coupled Logistic Maps Systems A and B demonstrate to have
suitable behaviour for most stringent applications, such as crypto. Although this could seem promising enough
for the algorithm, a further improvement could be achieved if one takes advantage of the specific symmetry
characteristics of these chaotic systems.

\begin{figure}[]
\centerline{\includegraphics[width=6.5cm]{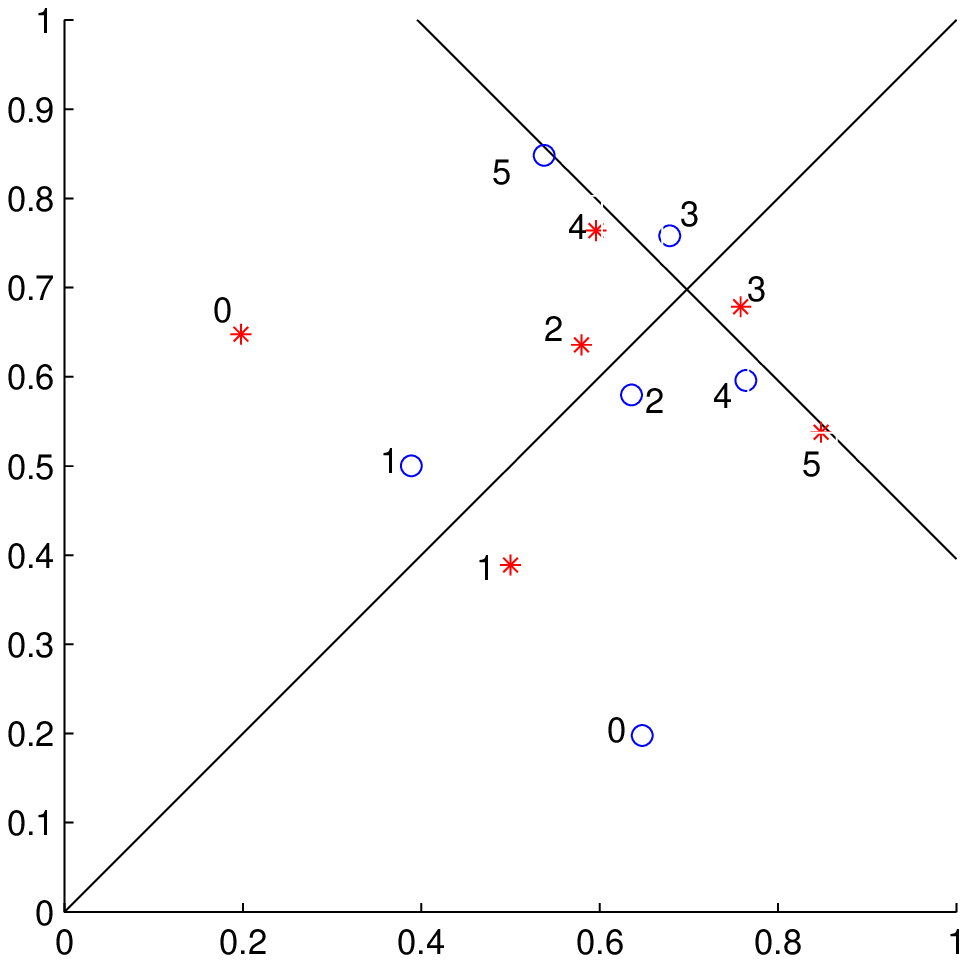}\hskip 0mm\includegraphics[width=6.5cm]{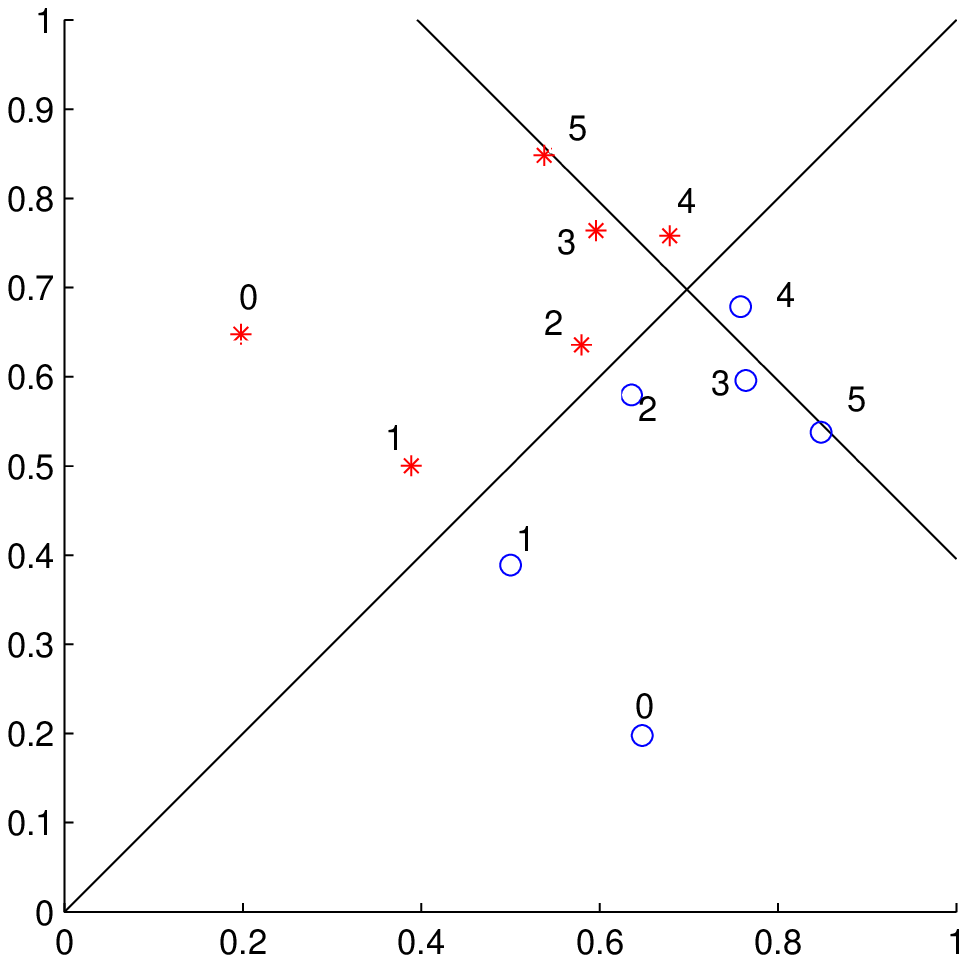}}
\centerline{(a)\hskip 6.5cm (b)}\caption{Five points in the evolution of trajectories for system A (a) or system
B (b) when starting with $\lambda=1.07$ and the same initial conditions $[x_0,y_0]$ (point $0$ with star marker)
and its coordinate conjugated $[y_0,x_0]$ (point $0$ with circle marker). It can be observed the symmetry of the
conjugated trajectories and the difference between systems A and B. In system A there is a jump along the
diagonal axis with every iteration.} 
\label{fig9}
\end{figure}

Let us observe that the systems under consideration present a symmetry with respect to the diagonal axis.
Consider now a simple interchange (or swap) of coordinates x and y in an orbit state. This produces a jump to a
conjugated orbit (see Fig. \ref{fig9}) but the attractor and the chaotic regime are not affected.

In these circumstances, a swapping of coordinates could be introduced in the algorithm of Fig. \ref{fig6},
without altering its pseudorandom properties. In practice, the swapping can be an additional step at the input
of the system, which is applied at specific instants $i-1$ as desired. When the swapping is applied at a
constant rate S, a swap of coordinates is introduced every S iterations, or instants of time. This means the
following performance: either of the systems evolves along one specific orbit during a number of S iterations,
then a swap in the coordinates is introduced (swapping $x \leftrightarrow y$) and a jump to a conjugated orbit
is produced.

Let us call S the rate of swapping or the swapping factor. In Fig. \ref{fig9} a schematic diagram is presented
to depicter more clearly the swapping procedure in Systems A and B. When starting with the same initial
conditions, one orbit and its conjugated are presented, jumping from one to the other is possible thanks to the
swapping factor.

This novel mechanism of swapping is named by the authors as \emph{symmetry-swap}, for it consists of a swap
between coordinates in mappings with particular symmetry characteristics. The interesting thing about it, is
that no matter what number of consecutive iterations and swaps are performed to the system, the chaotic
behaviour always prevails. Logically, this particular fact will make pseudo-randomness to prevail too.

\begin{figure}[]
\centerline{\includegraphics[width=5.8cm]{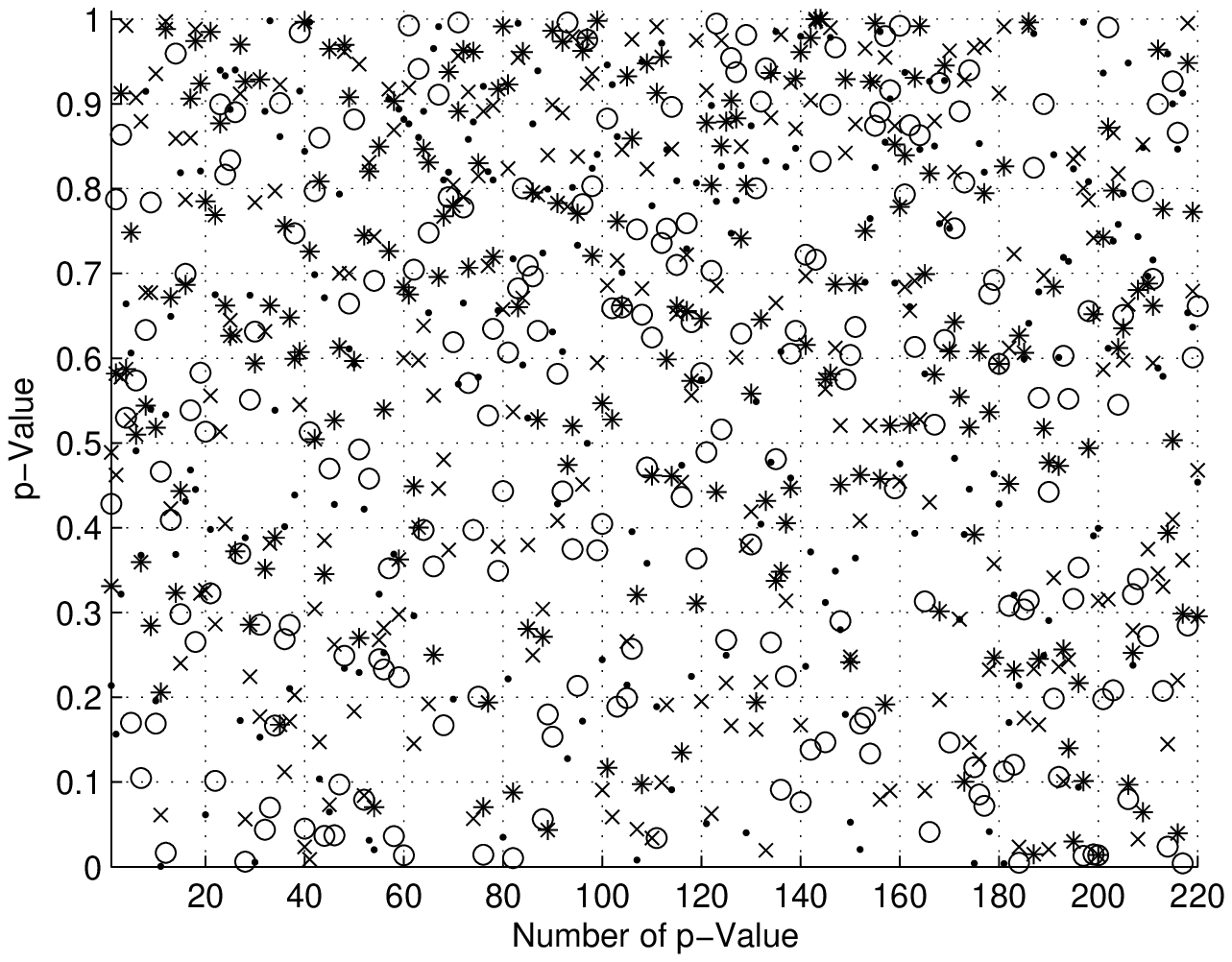}\hskip 5mm\includegraphics[width=5.8cm]{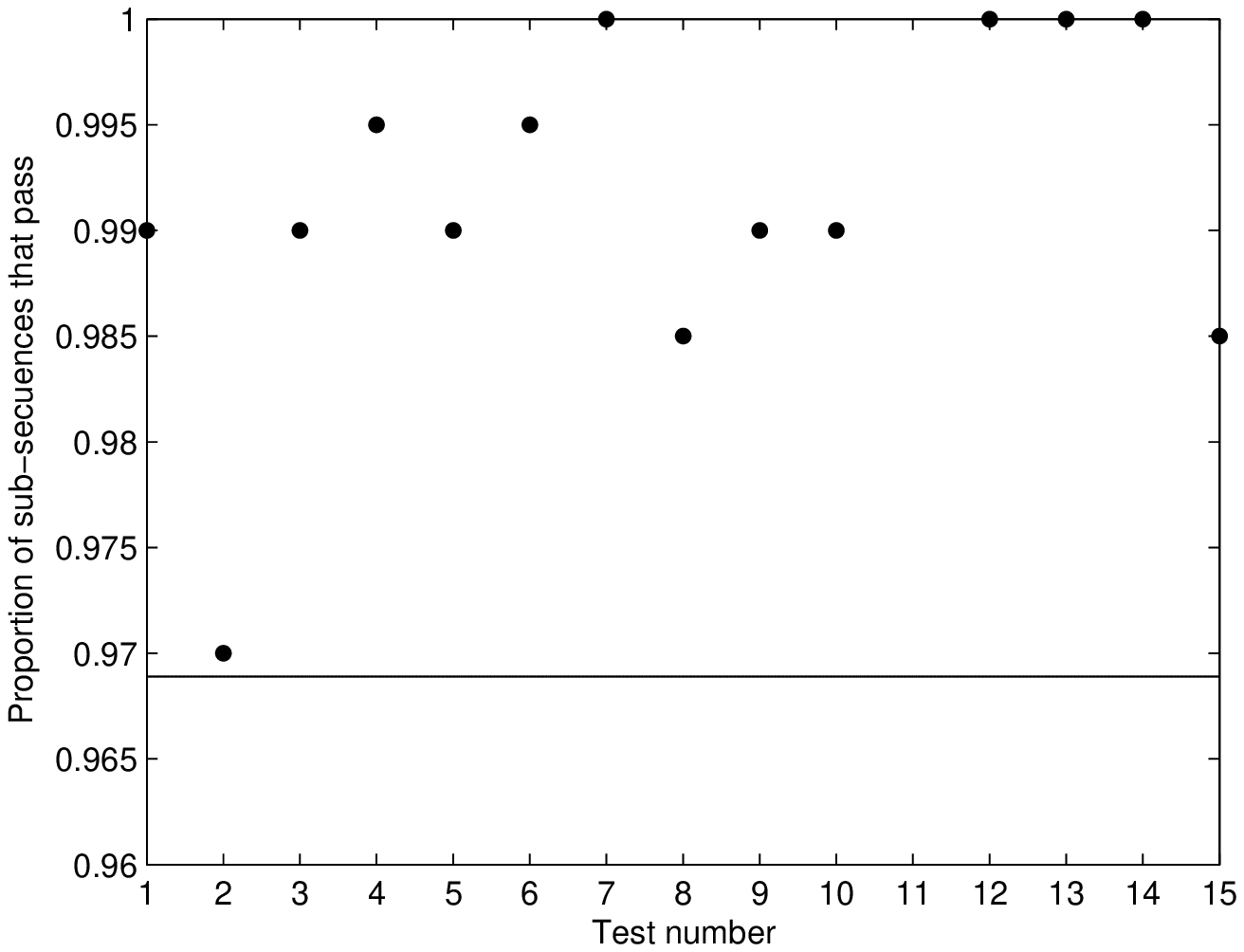}}
\centerline{(a)\hskip 5.5cm (b)} \centerline{\includegraphics[width=6cm]{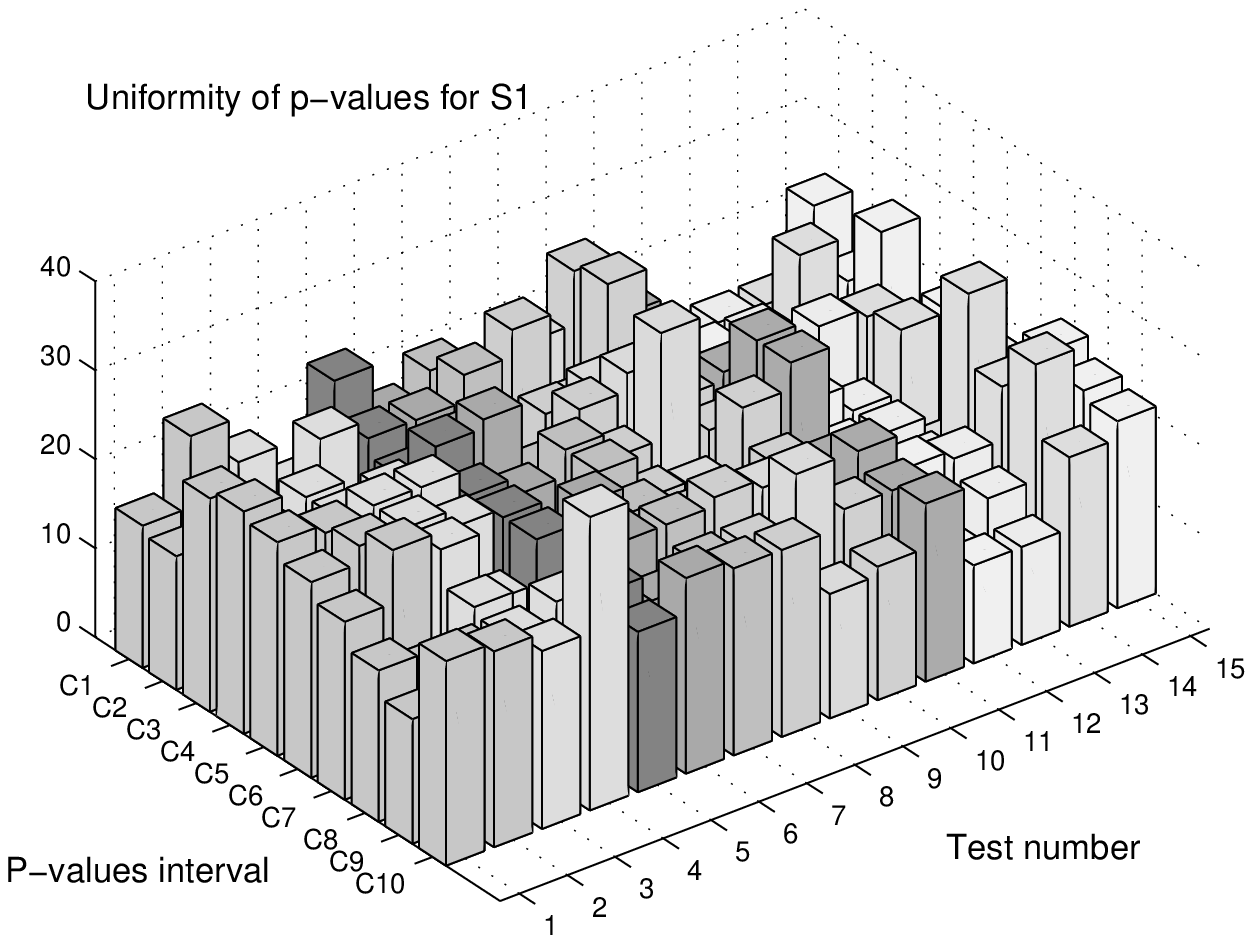}} \centerline{(c)} 
\caption{In (a), p-values obtained with all tests of Diehard test suite for initial conditions S1,S2,S3 and S4 with
$P=P_{Dmin}$ and $S=50$. In (b), it is displayed the proportion of sub-sequences that pass NIST tests suite with
initial conditions S1, $P=P_{Nmin}$ and $S=50$. In (c), the distribution of p-values for each test with the same
conditions of (b) demonstrates the required uniformity.}
 \label{fig10}
\end{figure}

The authors explored the construction of a swapped PRBG following algorithm in Fig. \ref{fig6} with system A as
described in section 5, and adding a constant swapping factor of value $S$ in the input. Ten pseudo-random
binary sequences were generated with the same characteristics (initial conditions and length) as the ones
described in Table 3. A swapping factor of $S=90$ was applied for sequences to be tested with Diehard test
suite. To illustrate a different value, a swapping factor of $S=50$ was chosen with NIST's suite. The sequences
of the swapped PRBGs demonstrated similar random results when submitted to the tests. Very similar $P_{min}$
values to those in Table 3, or even the same, were obtained in all cases. Fig. \ref{fig10} shows graphically the
results obtained and illustrated the success of the tests. Unsurprisingly, this demonstrates that in this case
the \emph{symmetry-swap} maintains pseudo-randomness.

It is important to observe at this point, that the introduction of a swapping factor $S$ does not penalizes the
computational cost of the resulting PRBG. Its asymptotic behaviour is again dominated by the chaotic block. As a
result, the swapped  2D symmetric coupled logistic maps PRBGs have an asymptotic tight bound of order
$\Theta(P*n)$.

Another valuable aspect to remark is, that the swapping factor $S$ can offer an improvement in the range of
input values of the initial PRBG algorithm. In cryptography, this means an enhancement in security and it can be
obtained straight from the fact that $S$, considered as a constant value, may represent a new free parameter in
the key-space.

Let us consider that the useful values of $S$ could range in the interval $S=[1,n]$, where $n$ is the number of
bits generated. Taking n for a typical value of 1 Mill. of bits, this would enlarge the key space calculated in
subsection 5.3. Following analogous calculations, with $[S]=1000000$ and $K_{S}=[S]/\epsilon$, then $K=
K_{\lambda}\times K_{x_0}\times K_{y_0}\times K_{P}\times K_{S}$ will be increased to 143 for single precision
and to 288 for double precision. The enlargement of the key space makes the swapped algorithm stronger against
brute force attack than the non-swapped one.

Even more, one may think that the introduction of a swapping factor $S$ can be applied in multiple ways.
Consider, for example, different values of $S$ used alternatively in the process, this may make the swapping
factor many dimensional. Another way could be to consider an $S$ value variable in time. The swapping factor can
also offer an easy feedback mechanism, when making its value dependable of the output. Therefore the
\emph{symmetry-swap} mechanism is a very flexible tool.

In the end, it can be observed that the \emph{symmetry-swap} offers a remarkable advantage, while maintaining
speed and simplicity of the initial PRBG algorithm.

\section{Conclusions}

In the present work, a refinement of the algorithm exposed in ~\cite{suneel} by M. Suneel is presented. It
consists of the introduction of a finite automata that makes possible its application to other chaotic maps. In
some way, this finite automata could be said to extend the range of application of this algorithm for other 2D
chaotic systems. This is referred in ~\cite{Li_thesis} as making the PRBG chaotic-system-free.

The fact is that, while systematic, the scheme presented in this paper is not straightforward. This is because
building the finite automata requires necessarily a detailed study of the geometrical properties of the
dynamical evolution of the chaotic system.

The authors apply this technique to build two new PRBG using two particular 2D dynamical systems formed by two
symmetrically coupled logistic maps. A set of different pseudo-random sequences are generated with one of the
PRBG. Statistical testing of these sequences shows fine results of random properties for the PRBG.

The estimation of the PRBG computational cost gives an asymptotic tight bound of $\Theta(P*n)$. The available
size of input values or the key space is also calculated and a minimum length of binary keys of 100 and 216 bits
is obtained for simple and double precision respectively. These preliminary results indicate a promising quality
of the PRBG for cryptographic applications.

Finally, an enhancement of the previous PRBGs is obtained exploiting the symmetry characteristic of the Coupled
2D Logistic maps. This is done by a new mechanism named as \emph{symmetry-swap}, that consists of a coordinate
swapping operation in the input variables of the chaotic systems. This gadget introduces an arbitrary change of
orbit in the evolution of the chaotic system. This novel strategy is only possible due to the symmetry inherent
and characteristic of the Coupled 2D mappings.

It is observed that the \emph{symmetry-swap} gives an additional degree of freedom to the chaotic PRBG algorithm
without additional computational penalties. After obtaining this enhanced or swapped PRBG, it is shown that the
computational cost and pseudo-random properties are similar to the previous PRBGs obtained with the non-swapped
algorithm. The input values or key space is, however, largely increased. Swapping represents a novel strategy
for finding additional degrees of freedom in the key space of a chaotic PRBG. Introducing the sampling factor
$P$ as an additional degree of freedom forces the designer to consider a trade-off between the range of $P$
values and the speed of the algorithm. On the contrary, introducing the swapping factor $S$ implies no extra
computational cost. Moreover this degree of freedom can be introduced in multiple ways. Some examples are to
consider it as a constant value, as a time varying one or as a feedback mechanism. Therefore the swapping factor
$S$ can increase the security of the system with great flexibility.

The role of geometry and symmetry properties in the chaotic PRBG algorithm presented here has been proved
noteworthy. This has been so, to the point that valuable achievements have been obtained from them. The authors
hope that similar considerations on other PRBGs may be useful and help in achieving comparable results.

{\bf Acknowledgements}
The authors acknowledge some financial support by spanish grant DGICYT-FIS200612781-C02-01.


\end{document}